\documentclass[9pt,twocolumn,twoside]{pnas-new}

\usepackage{hyperref}
\hypersetup{pdfauthor={Palazzi Saverio},pdftitle={},%
	colorlinks=true, linktocpage=true, pdfstartpage=3, pdfstartview=FitV,%
	breaklinks=true, pdfpagemode=UseNone, pageanchor=true, pdfpagemode=UseOutlines,%
	plainpages=false, bookmarksnumbered, bookmarksopen=true, bookmarksopenlevel=1,%
	hypertexnames=true, pdfhighlight=/O,%
	urlcolor=magenta, linkcolor=cyan, citecolor=red, 
}  

\usepackage{pifont}

\usepackage{pdfpages}

\pdfoutput=1 
\usepackage{graphicx}  
\usepackage{dcolumn}   
\usepackage{bm}        
\usepackage{amssymb}   
\usepackage{comment}
\usepackage{amsmath,amssymb,amsthm,dsfont,amsfonts,xfrac,braket,mathtools,bm,bbm,mathrsfs}

\usepackage{hhline}

\usepackage{subcaption}

\usepackage{makecell}

\usepackage{xcolor}

\usepackage{comment}

\usepackage[english]{babel}
\usepackage{amsopn}
\usepackage{float}

\usepackage{stackengine}

\newcommand{\sign}{\text{sign}}

\newcommand{\quotationmarks}[1]{``#1''}
\newcommand{\DUFC}{D_\text{uc}^\text{\tiny FC}}
\newcommand{\DU}{D_\text{uc}}

\usepackage{url}
\usepackage{enumitem}

\usepackage{graphicx}
\usepackage{dcolumn}
\usepackage{bm}
\usepackage{color}

\usepackage{float}

\usepackage{tikz}
\usepackage{scalerel}
\usepackage{pict2e}
\usepackage{tkz-euclide}
\usetikzlibrary{calc}
\usetikzlibrary{patterns,arrows.meta}
\usetikzlibrary{shadows}
\usetikzlibrary{external}

\templatetype{pnasresearcharticle} 

\begin{document}

\title{Critical exponents of the spin-glass transition in a field at zero temperature}

\author[a,b]{Maria Chiara Angelini}
\author[a,1]{Saverio Palazzi} 
\author[c,a,b,d,1]{Giorgio Parisi}
\author[e,a,b]{Tommaso Rizzo}

\affil[a]{Dipartimento di Fisica, Sapienza Università di Roma, P.le Aldo Moro 5, 00185 Rome, Italy}
\affil[b]{Istituto Nazionale di Fisica Nucleare, Sezione di Roma I, P.le A. Moro 5, 00185 Rome, Italy}
\affil[c]{International Research Center of Complexity Sciences, Hangzhou International Innovation Institute, Beihang University, Hangzhou 311115, China}
\affil[d]{Institute of Nanotechnology (NANOTEC) - CNR, Rome unit, P.le A. Moro 5, 00185 Rome, Italy}
\affil[e]{Institute of Complex Systems (ISC) - CNR, Rome unit, P.le A. Moro 5, 00185 Rome, Italy}

\leadauthor{Angelini}

\significancestatement{Spin glasses—systems with disordered magnetic interactions—serve as a fundamental model for studying complexity in physics. While their behavior is well characterized in fully connected (i.e., infinite-dimensional) systems, a comprehensive theory for real-world spin glasses in finite dimensions remains elusive. In particular, although the existence of a phase transition under a magnetic field is firmly established in infinite dimensions, its presence in finite dimensions remains controversial. We introduce an approach based on the Bethe approximation (an advanced mean-field framework) to demonstrate that spin glasses indeed undergo a transition into a frozen, disordered phase even in finite-dimensional systems. Furthermore, our method provides quantitative predictions for critical properties near this transition.}

\authorcontributions{All authors contributed equally to this work.}
\authordeclaration{The authors declare no conflict of interest.}
\correspondingauthor{\textsuperscript{1}To whom correspondence should be addressed. E-mail: giorgio.parisi@roma1.infn.it, saverio.palazzi.1997@gmail.com}


\begin{abstract}
We analyze the spin glass transition in a field in finite dimension $D$ below the upper critical dimension directly at zero temperature using a recently introduced perturbative loop expansion around the Bethe lattice solution. The expansion is generated by the so-called $M$-layer construction, and it has $1/M$ as the associated small parameter. Computing analytically and numerically these non-standard diagrams at first order in the $1/M$ expansion, we construct an $\epsilon$-expansion around the upper critical dimension $\DU=8$, with $\epsilon=\DU-D$. Following standard field theoretical methods, we can write a $\beta$ function, finding a new zero-temperature fixed-point associated with the spin glass transition in a field in dimensions $D<8$. We are also able to compute, at first order in the $\epsilon$-expansion, the three independent critical exponents characterizing the transition, plus the correction-to-scaling exponent.
\end{abstract}

\dates{This manuscript was compiled on \today}
\doi{\url{https://doi.org/10.1073/pnas.2511882122}}

\maketitle
\thispagestyle{firststyle}
\ifthenelse{\boolean{shortarticle}}{\ifthenelse{\boolean{singlecolumn}}{\abscontentformatted}{\abscontent}}{}

\firstpage[13]{5}

\dropcap{S}pin glasses (SG) with an external field are the prototype of disordered models.
Their fully connected (FC) mean-field (MF) version, the Sherrington and Kirkpatrick (SK) model \cite{sherrington1975solvable}, was solved more than forty years ago \cite{parisi1980sequence,parisi1980order}, and the solution has been proved to be rigorously correct more than twenty years later \cite{talagrand2003generalized,panchenko2013sherrington}. At small temperatures and fields, the SK model is in an SG phase, with an infinite number of equilibrium pure states, diverging susceptibilities, and frozen local magnetizations, while at high temperature/field it is in a paramagnetic (PM) phase.

Beyond MF, things are much less clear. In particular, it is still a debated question whether the SG model with an external field in finite dimensions admits an SG phase at low temperature, and which is the lower critical dimension.
The interpretation of numerical simulations is also debated, because of large finite-size effects and equilibration times \cite{baity2014three,baity2014dynamical,banos2012thermodynamic, vedula2023study, aguilar2024evidence, vedula2024evidence}.

A standard statistical mechanics tool for inferring the finite-dimensional behavior of models is the perturbative Renormalization Group (RG) \cite{Parisi1988,Amit_2006}. 
This approach has been applied to the spin glass model in a field and at finite temperature in several papers
\cite{bray1980renormalisation,temesvari2002generic,pimentel2002spin,moore2011disappearance,parisi2012replica,temesvari2017physical,charbonneau2017nontrivial,charbonneau2019morphology, holler2020one}.
For dimensions $D$ higher than the upper critical one $D>\DUFC=6$, the standard field-theoretical approach finds an MF-FC Fixed Point (FP) that is stable, even if its basin of attraction becomes smaller with decreasing dimension and eventually goes to zero at $\DUFC$.
For $D<\DUFC=6$, at the first order in the loop expansion, one cannot find a stable FP \cite{bray1980renormalisation, pimentel2002spin}. Going to the second order in the expansion \cite{charbonneau2017nontrivial,charbonneau2019morphology}, one finds a strong-coupling FP that could in principle be stable even above $\DUFC$, but in this strong-coupling regime, the perturbative nature of the expansion does not ensure to give correct results.

One could also use real-space RG methods, as done in ref.
\cite{gardner1984spin,parisi2001renormalization,drossel2000spin,angelini2013ensemble,angelini2015spin,monthus2015fractal,wang2018fractal}. 
However, they usually rely on some crude approximations and, even if they can provide useful indications, they are not conclusive.

In ref. \cite{Altieri_2017} a different perturbative loop expansion has been proposed, around the MF Bethe solution, a refined, finite-connectivity mean-field theory that is exact on the Bethe lattice (BL). The BL is a tree-like lattice in which the average connectivity is finite and the average length of loops diverges logarithmically with the size of the system. This property implies that, if a single pure state exists, nearest neighbors can be considered independent if one removes the link between them, and the marginals for each degree of freedom can thus be obtained by solving some self-consistent equations. 
SG models in a field can be solved on the BL exactly in the PM phase. They display a transition towards an SG phase at a small temperature/field which can be described within the so-called 1RSB approximation both at finite and zero temperature \cite{mezard2001bethe,mezard2003cavity,Parisi_2014}.
However, the BL solution is deeply different from the fully-connected one, in fact, the finite connectivity implies local fluctuations of the order parameter, that cannot be identified as a global averaged one: in this sense, the BL is much more similar to finite-dimensional systems.
The loop expansion of ref. \cite{Altieri_2017} is obtained through the following procedure: One creates $M$ copies of the original finite-dimensional lattice that at the beginning are thus independent. At this point a local random rewiring of the links is performed: we will call the obtained lattice the $M$-layer lattice.
More precisely, consider an edge in the original lattice and its $M-1$ copies. The random rewiring procedure involves detaching all $M$ edges (the original one and its copies) and reconnecting them uniformly at random, in such a way that each site must retain exactly one of the $M$ edges. Once this procedure is applied to every edge in the lattice, different copies (also called \quotationmarks{layers}) may become connected, while the connectivity of each site remains unchanged from that of the original lattice. For a pictorial representation see Fig. 1 of ref. \cite{angelini2020loop}.
In the large $M$ limit, the solution of the model will exactly correspond to the BL one, with topological loops whose length will diverge logarithmically with the number of total degrees of freedom. One could then perform an $1/M$ expansion that will take the form of a diagrammatic expansion in the number of topological loops with appropriate rules \cite{Altieri_2017}. Using standard RG methods, one could then identify the upper critical dimension of the model $\DU$, and set up an expansion around $\DU$ to obtain critical properties and in particular the critical exponents in lower dimensions. When this expansion is applied to models that have the same kind of transition both on the BL and on the FC lattice, one recovers exactly the same results as the standard field-theoretical loop-expansion, as recently shown for the Ising model \cite{Angelini_2024Ising}, for the percolation model \cite{angelini2025bethe} and for the SG in a field in the limit of high connectivity for $T>0$ \cite{angelini2018one}.
Instead, when the BL solution is different from the FC one (or if the FC solution does not exist), the $M$-layer expansion gives completely new results: this is the case for the Random Field Ising model (RFIM) at zero temperature \cite{angelini2020loop}, the bootstrap percolation \cite{rizzo2019fate}, the glass crossover \cite{rizzo2020solvable} and the Anderson localization \cite{baroni2024corrections}.

In ref. \cite{angelini2022unexpected} the $M$-layer expansion has been applied to the SG in a field directly at $T=0$: there is no transition in the SK model at $T=0$, because the system is in the SG phase no matter which is the value of the field, while in the BL there is a critical value of the field that divides the PM and the SG phases at $T=0$ and the loop-expansion is performed around this point. Excitingly the $M$-layer expansion around the BL found an upper critical dimension $\DU=8$, at variance with the upper critical dimension identified by standard field-theoretical analysis at $T\neq 0$ that is $\DUFC=6$ \cite{bray1980renormalisation}. In this paper, we proceed along the path started with ref. \cite{angelini2022unexpected}, and we set up an expansion below $\DU=8$: following standard RG methods, we can identify a 
perturbative finite-dimensional $T=0$ fixed-point that is stable for $D<\DU$, that governs the finite-dimensional SG transition. We are also able to write perturbative expressions for the independent critical exponents: as in standard $T=0$ transitions, the number of independent exponents is three and not two as in $T\neq0$ fixed-points \cite{Bray_1985}. We also carried out the same computation for the critical exponents of a one-dimensional SG model with long-range interactions, for which the power-law
exponent that controls the strength of the interactions can be linked to an effective dimension. The results of this computation are particularly useful when 
one wants to compare analytical results with numerical simulations, for which one-dimensional long-range models are much more suitable.

\section{Model and Observables}
\label{sec:model}

We consider the Edwards-Anderson SG model at zero temperature, defined on a generic lattice, by the following Hamiltonian:
\begin{equation}
    \mathcal{H}\left(\{\sigma_i\}_{i\in \mathcal{L}}\right)=-\sum_{(i,j)\in \mathcal{E}}J_{ij}\sigma_i\sigma_j -  \sum_{i \in \mathcal{L}} H_i \,  \sigma_i\,,
\label{eq:H}
\end{equation}
where $\mathcal{L}$ is the set of lattice sites, $\mathcal{E}$ is the set of edges and $\sigma_i=\pm1$ $\forall i \in \mathcal{L}$ are Ising spins at the lattice sites.
The $J_{ij}$'s and the $H_i$'s are quenched random variables, in particular, we will consider the cases of i) Gaussian distributed $J$'s with $\mathbb{E}_J[J_{ij}]=0$ and constant field $H_i=H\  \forall i \in \mathcal{L}$; and ii) Bimodal distributed $J_{ij}=\pm 1$ with zero means and Gaussian distributed random fields $H_i$'s with variance $\Delta^2$ and zero means. In the first case, the transition will occur at some critical value $H_c$ of the field while in the second case, it will occur at some critical value $\Delta_c$ of the variance.

On the Bethe lattice, the above model has a zero-temperature transition and we are interested in assessing the fate of this transition in finite-dimensional models. To do so we have considered the $M$-layer lattice. The aforementioned construction yields a random finite-dimensional lattice characterized by the number of layers $M$. In any dimension, when $M$ goes to infinity all observables of the model converge to the Bethe lattice result, therefore a phase transition is observed with the Bethe lattice critical exponents. At finite values of $M$ there are $\mathcal{O}(1/M)$ corrections that are harmless above the upper critical dimension $D>\DU=8$, but are expected to alter the critical behavior for $D<8$ \cite{angelini2022unexpected}. 
In the following, we will consider the first $1/M$ corrections to various observables and show that for $D<8$ they do not destroy the transition leading instead to an expansion of the critical exponents in powers of $\epsilon \equiv 8-D$. A similar conclusion will be found for the corresponding long-range models.

Given a realization of the disorder, we consider for each spin $\sigma_i$ the local field $h_i$, defined as usual such that i) $\sigma_i$ is oriented along $h_i$ in the ground state and ii) $\Delta E_i=2|h_i|$ is the \quotationmarks{excitation energy}, i.e. the energy difference between the ground state and the ground state where $\sigma_i$ is constrained in the direction opposite to $h_i$.
We stress that the local field $h_i$ is a complicated function of all $J_{ij}$'s and of the external fields $H_i$'s in the system.
The single-site probability density over the disorder that $h_i$ is equal to $h$ is given by $P_1(x_1,h)$ where $x_1$ is the position of $\sigma_i$. 
Due to the translational invariance of the disorder, the distribution is the same for all spins i.e. $P_1(x_1,h)=P_1(h)$. 

Since either the $J_{ij}$'s or the $H_i$'s obey continuous distributions, $P_1(h)$ is also continuous, therefore one could expect the probability of finding two spins at a finite distance with the exact same local field to be zero. It turns out that this is not the case and that for each value of $\Delta E$ there is a finite probability density over the disorder of finding a cluster of spins with $|h_i|=\Delta E/2$. 
Therefore, for a given realization of the disorder, the whole system is partitioned in clusters of spins with the same excitation energy. In the PM phase, each cluster contains a finite number of sites and extends over a finite region of the lattice. The clusters are related to the concept of avalanches, indeed one can show that, if a spin is forced in the opposite direction of its local field, all the spins in the same cluster are also flipped in the new ground state. For the sake of readability we postpone to Sec. (\ref{sec:cluseava}) a detailed discussion of clusters and avalanches. The clusters with nearly zero excitation energy $\Delta E=0^+$ are of particular importance at the critical point: we call them \quotationmarks{soft} clusters as they can be flipped with zero energy cost. We will see that approaching the de Almeida-Thouless line \cite{deAlmeida_1978}, the typical size of the soft clusters diverges, much as it happens in percolation when the critical occupation probability is approached from the non-percolation phase \cite{Stauffer94,coniglio2000geometrical}.

The notion of clusters allows us to straightforwardly define correlation functions. A generic multi-point correlation function $\mathcal{P}_q(x_1,\dots,x_q;\Delta E)$ is defined as the probability density over the disorder that the $q$ spins at positions $x_1,\dots,x_q$  are in the same cluster with excitation energy $\Delta E$. Again, as the disorder distribution is translational invariant, the correlation functions are translational invariant as well. In particular, the probability $\mathcal{P}_1(x_1;\Delta E)$ that the spin at position $x_1$ belongs to a cluster with excitation energy $\Delta E$ does not depend on $x_1$ and is simply related to the local field distribution: $\mathcal{P}_1(x_1;\Delta E) =P_1(\Delta E/2)/2+P_1(-\Delta E/2)/2$.

Guided by the percolation problem we also consider the statistics of cluster sizes. We define the cluster density $n(s,\Delta E)$ such that, in a system of size $N$, the number of clusters of size $s$ and excitation energy between $\Delta E$ and $\Delta E+dE$ is given by $n(s,\Delta E) N dE$. In principle, $n(s,\Delta E)$ depends on the given realization of the disorder but, on general grounds, we expect it to be self-averaging at large $N$.
Much as in percolation, one sees that the correlation functions are related to the moments of $n(s,\Delta E)$, in particular, we have 
\begin{equation}
  \sum_s  \, s \, n(s,\Delta E) = \mathcal{P}_1(x_1;\Delta E) \ . 
\end{equation}
Furthermore, the sum of $ \mathcal{P}_{2}(x_1,x_2;\Delta E)$ over $x_2$ is related to the second moment of the cluster distribution:
\begin{equation}
  \sum_s  \, s^2 \, n(s,\Delta E) = \sum_{x_2} \mathcal{P}_{2}(x_1,x_2;\Delta E) \, 
\label{eq:s2}
\end{equation}
and in full generality:
\begin{equation}
  \sum_s  \, s^q \, n(s,\Delta E) = \sum_{x_2,\dots,x_q} \mathcal{P}_q(x_1,\dots,x_q;\Delta E) \, . 
\label{eq:s^qP_q}
\end{equation}
Note that the RHS of the last three formulas does not depend on $x_1$ due to translational invariance.

The notion of local fields $h_i$ can be generalized to any number of spins. In particular, given two spins $\sigma_i$ and $\sigma_j$ and a specific instance of the disorder, we consider the effective energy function $E_{ij}(\sigma_i,\sigma_j)=-u_i \sigma_i-u_j \sigma_j - J_{ij}^{eff} \, \sigma_i \sigma_j$ that yields, apart from a constant, the ground state energy of the system when the two spins are constrained in the four possible configurations $\sigma_i=\pm 1$, $\sigma_j=\pm 1$. Note, again, that the effective fields $u_i$, $u_j$ and $J_{ij}^{eff}$ are complicated functions of the actual couplings  $J$'s and fields $H$'s of the whole system, as given in \eqref{eq:H}. 
We thus introduce the probability density over the disorder that the triplet $(u_i,u_j,J_{ij}^{eff})$ is equal to  $(u,u',J)$ as $P_2(x_1,x_2; u,u',J)$ where $x_1$ and $x_2$ denote respectively the position on the lattice of the sites $i$ and $j$. 
Again, this distribution is translational invariant with respect to a shift of $x_1$ and $x_2$, besides it is symmetric with respect to the exchange $u \leftrightarrow u'$. 
We expect that at large distances $|x_1-x_2|$ this distribution converges to the factorized form $P_1(u)P_1(u')\delta(J)$ where $P_1(u)$ is the probability distribution of the local field on a given site, indeed if $J=J_{ij}^{eff}=0$, then $u_i=u$($u_j=u'$) coincides with the local field $h_i$($h_j$).   
An important property of the triplets distribution is that, at any distance $|x_1-x_2|$, there is a finite probability of having $J=0$, see for instance Table S1 in the Supplementary Information.
Therefore we can define two additional distributions $P^{dis}(x_1,x_2;u,u')$ and $P^{con}(x_1,x_2;u,u',J)$, that take into account triplets with respectively $J=0$ and $|J|>0$:
\begin{multline}
P_2(x_1,x_2;u,u',J)-P_1(u)P_1(u')\,\delta(J) = \\ P^{dis}(x_1,x_2;u,u')\,\delta (J)+ P^{con}(x_1,x_2;u,u',J) \, ,   
\label{eq:Pdiscon}
\end{multline}
where $P^{con}(x_1,x_2,u,u',J)$ is regular at $J=0$. Note that triplets with strictly vanishing $J$ are also present in the RFIM, whose critical behavior is described by a zero temperature fixed point.

In the following, we will study the probability densities $P_1$, $P_2$, and $\mathcal{P}_q$ on the $M$-layer random lattice.
In this case, the average over the disorder includes the average over the rewirings, i.e. over all possible realizations of the $M$-layer. On the $M$-layer lattice a point is specified by its position $x$ on the original lattice and by its layer. However, after averaging over all possible realizations of the $M$-layer, the probabilities  $P_1$, $P_2$ and $\mathcal{P}_q$ do not depend on the actual layers of the spins but only on their position on the original lattice, therefore, the dependence on the layers will be dropped.

\subsection{Scaling Laws}\label{sec:scaling_laws}

An explicit computation, done in the SI Appendix, shows that at leading order in the $1/M$ expansion \eqref{eq:Pdiscon} takes the following simple form after Fourier transform ($FT$) in position space: 
\begin{multline}
   FT\left[ P_2(x,x';u,u',J)-P_1(u)P_1(u')\delta(J)\right] \propto (2 \pi)^D \delta(k+k')\\
    \widehat{P}_1(u) \widehat{P}_1(u')\left(- \frac{1}{(k^2+t)^2} \delta(J) + \frac{1}{(k^2+t+|J|)^3}\right)
\label{eq:FTP}
\end{multline}
where $t$ vanishes linearly at the critical point i.e. $t \propto H-H_c$ or  $t \propto \Delta-\Delta_c$. 
The function $\widehat{P}_1(u)$ depends on the microscopic details of the model and satisfies i) $\widehat{P}_1(u) \geq 0$ for all $u$ and ii) $\widehat{P}_1(u)=\widehat{P}_1(-u)$ in the random-field case \footnote{Note that, applying a random transformation $\sigma_i \to -\sigma_i$ with probability $1/2$ independently for each spin, the constant field case reduces to the random field case with $H_i = \pm H$.}. As a consequence of $\widehat{P}_1(u)=\widehat{P}_1(-u)$, the average over the disorder of $\sigma_i \sigma_j$  on the ground state vanishes. This has to be contrasted with the RFIM where $\sigma_i \sigma_j$  on the ground state is always positive due to the presence of an additional term  $(2 \pi)^D \delta(k+k')
    A(u) A(u')(k^2+t)^{-2} \delta(J)$ in \eqref{eq:FTP} \cite{angelini2020loop} where $A(u)$ is antisymmetric.

Given a generic translational invariant two-point function $C(x,x')$, we define the associated correlation length as
$\xi^2 \equiv \int d^D\, x \, |x|^2 \, C(x,0)\big/\int d^D\, x  \, C(x,0)$. Thus in \eqref{eq:FTP}
two correlation lengths $\xi_{dis}$ and $\xi_{con}$ can be identified, respectively associated with the disconnected and connected functions.  
$\xi_{dis}$ diverges as $t^{-1/2}$, while $\xi_{con} \propto (t+|J|)^{-1/2}$ depends on both $t$ and $J$ (but not on $u$ and $u'$) and diverges iff both $t$ and $J$ vanish.
We note that \eqref{eq:FTP} holds at small momenta, that is at large distances $|x-x'|$, of the order of the correlation lengths that are large close to the critical point $(t=0)$ when $|J|\ll 1$. 

As we mentioned before, higher-order corrections in $1/M$ do not change \eqref{eq:FTP} above the upper critical dimension $\DU=8$ \cite{angelini2022unexpected}. For $D<8$, corrections are instead important, nonetheless, we expect that \eqref{eq:FTP} is generalized by the following scaling expression:
\begin{multline}
    P_2(x_1,x_2;u,u',J)-P_1(u)P_1(u')\delta(J)=\widehat{P}_1(u) \widehat{P}_1(u') \times\\
    \left(- \frac{1}{r^{D-4+\overline{\eta}}} f_{dis}\left( \frac{r}{\xi_{dis}}\right) \delta(J) +\frac{1}{r^{D-4+\overline{\eta} -\theta}} f_{con}\left( \frac{r}{\xi_{con}}, \frac{|J|}{t^{\nu \theta}}\right) \right)\,,
\label{eq:Pscaling}
\end{multline}
where $r \equiv |x_1-x_2|$ and again the length $\xi_{dis}$ depends only on $t$ and diverges as $\xi_{dis} \propto t^{-\nu}$ while $\xi_{con}$  depends on both $t$ and $J$ (but not on $u$ and $u'$) and obeys $\xi_{con}=t^{-\nu}f_{\xi}(|J|/t^{\nu \theta})$. 
 The scaling function $f_{\xi}$ is such that $\xi_{con}$ is finite unless both  $t$ and $J$ vanish and $\xi_{con} \propto t^{-\nu}$ for $J=0$ and $\xi_{con} \propto |J|^{-1/\theta}$ for $t=0$.
For $D<\DU$ the critical exponents ($\theta$, $\nu$, $\overline{\eta}$) and the scaling functions ($f_{dis}(\varrho)$, $f_{con}(\varrho,\jmath)$ and $f_{\xi}(\jmath)$) depend on the dimension but not on the microscopic details of the model, i.e. they are universal.
The two scaling functions $f_{dis}(\varrho)$ and $f_{con}(\varrho,\jmath)$ are finite for $\varrho=0$ and decrease exponentially for $\varrho \to \infty$.
The scaling variable $\jmath$ in  $f_{con}(\varrho,\jmath)$ and $f_{\xi}(\jmath)$ defines the lines of  approach of the point $(t=0,J=0)$ and goes from $\jmath=0$, corresponding to $J=0$, to $\jmath=\infty$ corresponding to the line $t=0$.
Note that \eqref{eq:FTP}  is a special case of \eqref{eq:Pscaling} with  $\theta=2$, $\nu=1/2$, $\overline{\eta}=0$ and appropriate scaling functions $f_{con}(\varrho,\jmath)$, $f_{dis}(\varrho)$ and $f_{\xi}(\jmath)$. 

 Much as in \eqref{eq:FTP} the function $\widehat{P}_1(u)$ is positive definite and model-dependent (at variance with $f_{dis}$ and $f_{con}$ that are universal). Given that, by definition, i) $P_2(x_1,x_2;u,u',J)$ is normalized with respect to $(u,u',J)$ and ii) $P_1(u)$ is normalized with respect to $u$, it follows that the integral over $(u,u',J)$ of the LHS of \eqref{eq:Pscaling} vanishes. On the other hand, since $\widehat{P}_1(u)$ on the RHS is positive definite, the RHS must vanish due to the integration over $J$. This has the following implications: i) the same exponent $\overline{\eta}$ appears in the connected and disconnected part, ii) $f_{dis}(\varrho)$ is related to $f_{con}(\varrho,\jmath)$ and $f_{\xi}(\jmath)$ by the following relationship: 
\begin{equation}
f_{dis}\left(\frac{r}{\xi_{dis}(t)}\right)=\int_{-\infty}^{\infty} f_{con}\left( \frac{r}{\xi_{con}(t,J)}, \frac{|J|}{t^{\nu \theta}}\right)\,dJ  \ ,
\label{cond1}
\end{equation}
iii) the ratio $\xi_{dis}(t)/\xi_{con}(t,J)$ depends solely on the ratio $|J|/t^{\nu \theta}$ and is universal.

The connected correlation function $\mathcal{P}_2(x_1,x_2;\Delta E)$ can be obtained from the distribution $P_2(x_1,x_2;u,u',J)$: as shown in the SI Appendix, we have to integrate $P_2(x_1,x_2;u,u',J)$ over $u$, $u'$ and $J$ with the conditions that $|u|<|J|$, $|u'|<|J|$, and $|u+u'\,\sign J|=\Delta E/2$. 
By means of a simple computation  we obtain from \eqref{eq:FTP} the following expression valid for $D \geq \DU$:
\begin{equation}
   FT[\mathcal{P}_2(x_1,x_2;\Delta E)] \propto
   (2 \pi)^D \delta(k+k')
    \widehat{P}_1^2(0) \frac{1}{k^2+t+\Delta E/4} \ .
\label{eqetaMF}
\end{equation}
Note that the associated correlation length is $\xi=(t+\Delta E/4)^{-1/2}$ and therefore diverges only for $\Delta E=0$ and $t=0$, i.e. only the soft clusters display critical behavior. For $D < \DU$ this generalizes to the following scaling expression valid at large $r$
\begin{equation}
    \mathcal{P}_2(x_1,x_2;\Delta E) =
    \widehat{P}_1^2(0)\, \frac{1}{r^{D-4+\overline{\eta}+\theta}}  \, f_2\left( \frac{r}{\xi}, \frac{\Delta E}{t^{\nu \, \theta}}\right)
\label{eqeta2}
\end{equation}
where $f_2(r/\xi,\Delta E/t^{\nu\theta})$ is a scaling function, related to $f_{con}$ and $f_{\xi}$ as shown in the SI Appendix, that decreases exponentially for $r/\xi \to \infty$ and is finite for $r/\xi\to 0$.  The correlation length $\xi$ depends on both $t$ and $\Delta E$; similarly to $\xi_{con}$, it obeys a scaling form $\xi=t^{-\nu}\tilde{f}_\xi(\Delta E/t^{\nu \theta})$ and diverges as $\xi \propto t^{-\nu}$ for $\Delta E=0$ and as $\xi \propto \Delta E^{-1/\theta}$ for $t=0$ \footnote{Note that the universal scaling function of $\xi_{con}(t,J)$ is not the same as that of $\xi(t,\Delta E)$.}. 
Defining the exponent $\eta$ from $\mathcal{P}_2(x_1,x_2,\Delta E) \propto 1/r^{D-2+\eta}$ we have 
\begin{equation}
    \eta= \overline{\eta}-2+\theta \ 
\label{eq:eta}
\end{equation}
and for $D \geq 8$ the values $\theta=2$, $\overline{\eta}=0$  imply the usual result $\eta=0$.
Then we see that, approaching the critical point $(t=0,\Delta E=0)$, the space integral diverges with $\xi$  as 
\begin{equation}
 \sum_{x_2}  \mathcal{P}_{2}(x_1,x_2;\Delta E) \,  \propto \xi^{2-\eta}  \, \tilde{f}_2\left( \frac{\Delta E}{t^{\nu\,\theta}}\right)
\label{eq:X2_scal}
\end{equation}
where $\tilde{f}_2(e)$ is related to $f_2(\varrho,e)$  (see the SI Appendix).
The corresponding expression for $D \geq \DU$ is given by \eqref{eqetaMF} evaluated for $k=0$ leading to
\begin{equation}
 \sum_{x_2}  \mathcal{P}_{2}(x_1,x_2;\Delta E) \,  \propto \frac{1}{t+\Delta E/4} \propto \xi^2
\end{equation}

The above result together with \eqref{eq:s2} suggests that $n(s,\Delta E)$ obeys a scaling law as well.
Following the analogous treatment of percolation \cite{coniglio2000geometrical,angelini2025bethe}  we assume that, changing the lengths by a factor $b$ by means of a real-space Renormalization Group (RG) transformation, the size $s$ of large clusters, the value of the typical energy excitation energy $\Delta E$ and the reduced field $t$ change according to:
\begin{equation} 
  s'=s/b^{D_f}\, , \ \Delta E' = \Delta E/b^{-\theta}\, , \ t'=t/b^{-1/\nu} \, , \ \xi' = \xi/b\ .  
\end{equation}
$D_f \leq D$ is by definition the fractal dimension of the clusters: it relates the linear size $l$ of a large cluster with its size $s$ using $s \propto l^{D_f}$. The fact that the exponent of the energy has to be identified with $\theta$ will be clear in the following.
Now we assume that under the RG transformation the cluster number at large $s$ is conserved, i.e.
\begin{equation}
   n(s,\Delta E,t) \Delta s \, \Delta (\Delta E) L^D \approx  n(s',\Delta E',t') \Delta s' \, \Delta  (\Delta E') (L')^D
\label{hyphyp}
\end{equation}
where $L$ is the linear size of the systems, $L^D=N$. 
Setting $b=(s/S)^{1/D_f}$ for some fixed large reference size $S$ we then obtain :
\begin{equation*}
    n(s,\Delta E,t) = s^{\frac{\theta-D-D_f}{D_f}} n\left(S, \frac{\Delta E}{ (s/S)^{-\theta/D_f}},\frac{t}{(s/S)^{-1/(\nu D_f)}}\right)
\end{equation*}
that can be rewritten in terms of $\xi(t,\Delta E)$ (the correlation length of $\mathcal{P}_{2}(x_1,x_2;\Delta E)$) as 
\begin{equation}
    n(s,\Delta E,t) \equiv s^{(\theta-D-D_f)/D_f} f_0(s / \xi^{D_f}, \Delta E / t^{\nu \theta})\ .
    \label{nlowd}
\end{equation}
$f_0(s,e)$ is a scaling expression that decreases exponentially at large values of $s$ and has a finite limit for $s=0$.
Therefore  the cluster density at fixed $\Delta E$ and $t$ follows a power-law $s^{(\theta-D-D_f)/D_f}$ up to a large value $s^*$ that diverges as $\xi^{D_f}$ approaching the critical point $t=0$ when $\Delta E\sim 0$.
We remark that the clusters with a finite $\Delta E>0$ are characterized by a finite correlation length including at $t=0$. Indeed we have $s^* \sim t^{-\nu \, D_f}$ at $\Delta E=0$ and $s^* \sim \Delta E^{-D_f/\theta}$ at $t=0$.
From \eqref{eq:s^qP_q} and \eqref{nlowd} we obtain:
\begin{multline}
 \sum_{x_2,\dots,x_q} \mathcal{P}_q(x_1,\dots,x_q;\Delta E) = \\
 \sum_s  \, s^q \, n(s,\Delta E)  \equiv \xi^{D_f q-D + \theta}\tilde{f}_{q}(\Delta E/t^{\nu \theta})\ . 
 \label{eq:def_K-point_function}
\end{multline}
In the following, we define the susceptibilities as the above integrals for $\Delta E=0$. They diverge at the critical point as:
\begin{equation}
    \chi_q \equiv \sum_{x_2,\dots,x_q} \mathcal{P}_q(x_1,\dots,x_q;0) \propto \xi^{D_f q -D+ \theta} \propto t^{-\nu (D_f q-D + \theta)}\ .
    \label{eq:def_K-point_susceptibility}
\end{equation}
The comparison of  \eqref{eq:def_K-point_function} for $q=2$ with \eqref{eq:X2_scal} leads to the identification of the energy exponent with $\theta$ and to:
\begin{equation}
    D_f=\frac{D-\theta+2-\eta}{2}\, ,\ \ \chi_q \propto \xi^{\left(\frac{q}{2}-1\right)(D-\theta)+\frac{q}{2}(2-\eta)}\ .
\label{eq:X2scalb}
\end{equation}
Note that this leads to $D_f=4$ for $D=\DU$.
Analogously to the connected susceptibility $\chi_2$ we introduce the disconnected susceptibility as:
\begin{multline}
   \chi_2^{dis} \equiv \sum_{x_2} P^{dis}(x_1,x_2;0,0) \propto \\
   -  \int_0^\infty \frac{1}{r^{D-4+\overline{\eta}}} \,f_{dis}\left( \frac{r}{\xi_{dis}}\right)  r^{D-1} dr \propto -\xi_{dis}^{4-\overline{\eta}} \propto -t^{4 \nu-\overline{\eta}\nu}  \ .
   \label{eqeta1}
\end{multline}
Thus we have $\chi_3^2/\chi_2^3 \propto t^{\nu( D-\theta)}$, while from Eqs. (\ref{eqeta1}) and (\ref{eqeta2}) we find $\chi^{dis}_2/\chi_2 \propto t^{\nu \theta}$. If the previous scaling laws hold, the quantity 
\begin{equation}
  \lambda \equiv  -\xi^{-D}\frac{\chi^{dis}_2 \, \chi_3^2}{\chi_2^4}
\label{def:lambda}
\end{equation}
should then remain finite at the critical point where $\xi$ diverges. We remark that in \eqref{eq:def_K-point_susceptibility} and in \eqref{def:lambda} $\xi$ stands for $\xi(t,0)$, i.e. the correlation length at $\Delta=0$ . Adopting standard jargon we will call $\lambda$ the renormalized coupling constant \cite{Parisi1988,Bellac_1991,Amit_2006} \footnote{Note that we put a minus sign in the definition of $\lambda$ to have a positive value at the critical point since $\chi_2^{dis}$ is negative}. 

An explicit computation (details elsewhere) shows that at leading order in $1/M$ the susceptibilities diverge as $\chi_q \propto \xi^{4 q -6}$, for $D \geq 8$ this is the correct result since higher order corrections in $1/M$ are irrelevant.
This implies that $\xi^{D_f q -D+ \theta}$ as given in \eqref{eq:def_K-point_susceptibility} is wrong and this can be traced back to the failure of \eqref{hyphyp} \footnote{This can be rationalized employing percolation theory. Within percolation, a box of size $b$ after a real-space RG transformation is considered occupied if there is a spanning cluster inside it \cite{Stauffer94}. 
Below the upper critical dimension, there is at most one spanning cluster in such a large box therefore the cluster number is conserved under the RG transformation. Instead, above the upper critical dimension there can be more than one spanning cluster in a large box of size $b$ \cite{coniglio2000geometrical} and therefore Eq. (\ref{hyphyp}) is incorrect.}. 
Indeed, in analogy with percolation \cite{coniglio2000geometrical,angelini2025bethe} we expect that for all $D>8$ the (hyper-scaling) cluster number expression (\ref{nlowd})  must be replaced with the $D=8$ result ($\eta=\overline{\eta}=0$, $\nu=1/2$,  $\theta=2$, $D_f=4$):
\begin{equation}
    n(s,\Delta E) = s^{-5/2} f_0(s / \xi^4, \Delta E / t) \ .
\label{nhighd}
\end{equation}
For $\Delta E=0$ we have indeed verified that the above expression gives $n(s,0)$ at leading order in $1/M$ leading to the aforementioned result $\chi_q \propto \xi^{4 q -6}$. Note that the resulting expression is precisely the same as the cluster number $n(s)$ in percolation above the upper critical dimension $D = 6$, which has a $s^{-5/2}$ tail with a cut-off that diverges as $t^{-2}$ and fractal dimension $D_f=4$ \cite{coniglio2000geometrical,angelini2025bethe}.

From \eqref{nhighd} it follows that $\chi_3^2/\chi_2^3 \propto \xi^{D-\theta}$ must be replaced by $\chi_3^2/\chi_2^3 \propto \xi^{6}$  and $\chi^{dis}_2/\chi_2 \propto \xi^{\theta}$ must be replaced by $\chi^{dis}_2/\chi_2 \propto \xi^{2}$ for $D>8$, meaning that, above the upper critical dimension, the renormalized coupling constant \eqref{def:lambda} goes to zero at the critical point as $\lambda \propto \xi^{8-D}$.

\subsection{Critical exponents for short-range interactions}

In the Materials and Methods section, we will show how the scaling laws can be used to transform the $1/M$ expansions into $\epsilon$ expansions for the critical exponents. In practice, the success of the procedure relies on the existence of a non-trivial zero of an appropriate \quotationmarks{beta-function} for $D<8$ and this provides, a posteriori, a validation of the scaling laws. The final result is:
\begin{multline}\label{eq:short_range_exponents}
    \nu=\frac{1}{2}+\frac{1}{16}\epsilon+\mathcal{O}(\epsilon^2)\, , \ \ \eta=-\frac{1}{16}\epsilon+\mathcal{O} (\epsilon^2)\, , \ \ \overline{\eta}=\frac{3}{16}\epsilon+\mathcal{O}(\epsilon^2) \, ,\\
    \qquad\qquad \omega=- \epsilon + \mathcal{O}(\epsilon^2)\, , \ \ \  \epsilon \equiv 8-D  \,. \hspace{2cm} \ 
\end{multline}
The exponent $\omega$ controls finite-size corrections: in a system of finite linear size $L$, a critical observable will display small corrections of order $L^\omega$.
\eqref{eq:eta} implies $\theta=2-\epsilon/4+\mathcal{O}(\epsilon^2)$.

\subsection{Long-range interactions}
We also carried out the computation of the critical exponents for a one-dimensional lattice with long-range (LR) interactions. In particular, we considered the spin glass model defined by the following Hamiltonian
\begin{equation}
    \mathcal{H}\left(\{\sigma_i\}_{i\in \mathcal{L}_{\text{LR}}}\right)=-\sum_{( i,j)\in\mathcal{E}_{\text{LR}}}J_{i j}\sigma_i\sigma_j - \sum_{i\in\mathcal{L}_{\text{LR}}} H_i\sigma_i\,,
\end{equation}
where the first sum is over all the edges $\mathcal{E}_{\text{LR}}$ of the LR system and $\mathcal{L}_{\text{LR}}$ is the corresponding set of sites. The couplings are long-range, meaning that $J_{ij}$ are independent distributed random variables that take non-zero values with a probability that decreases as a power-law function of $r_{ij}$, the distance between sites $i,j \in \mathcal{L}_{\text{LR}}$ \cite{Leuzzi_2008,Leuzzi_2009,Leuzzi_2011,Martin-Mayor_2012,Leuzzi_2013}. Specifically:
\begin{equation}
    \mathbb{P}[J_{ij}\neq0] \propto \frac{1}{r_{ij}^\rho}\,,
\end{equation}
where $1 < \rho < 3$. Once the non-zero coupling is extracted, its actual value is drawn from a Gaussian or bimodal distribution. As discussed in ref. \cite{Leuzzi_2008}, this diluted version of the problem is expected to lie in the same universality class as the fully connected model with Gaussian couplings with zero mean and variance decaying with the distance between sites \cite{Kotliar_1983}.
The zero-temperature critical phenomenology of the model is the same as the short-range (SR) models of Sec. (\ref{sec:model}). In particular,
\eqref{eq:Pscaling} holds with the factor $D-4$ replaced by $3-2 \rho$ and the $\eta$ exponent is defined from  $\mathcal{P}_2(x_1,x_2,\Delta E) \propto 1/r^{2-\rho+\eta}$.
In the limit $M \to \infty$ we find $\eta=\overline{\eta}=0$, $\theta=\rho-1$ and $\nu=(\rho-1)^{-1}$. These exponents are not changed by $1/M$  corrections for $1<\rho<\rho_{uc}=5/4$. Instead, for $\rho_{uc} \leq \rho \leq \rho_{lc}$, corrections are important.
The critical behavior of the susceptibilities is:
\begin{equation}
 \chi_2 \propto \xi^{\rho-1} \, , \   \chi_2^{dis} \propto \xi^{2(\rho-1)-\overline{\eta}} \, , \ \chi_q \propto \xi^{\rho-2-\overline{\eta}+\frac{q}{2}(1+\overline{\eta})}.
\end{equation}
As in the short-range case the $1/M$ expansion can be used \cite{angelini2026long} to extract the critical exponents for $5/4 \leq \rho \leq \rho_{lc}$ leading to:
\begin{multline}\label{eq:long_range_exponents}
    \nu=4+\mathcal{O}(\epsilon^2)\, , \ \ \ \ \ \eta=0\, , \ \ \ \ \ \ \ \overline{\eta}=\epsilon +\mathcal{O}(\epsilon^2)\, ,\\
    \omega=- 4\,\epsilon + \mathcal{O}(\epsilon^2) \, , \ \ \ \ \ \ \ \epsilon \equiv \rho-\frac{5}{4}  \,. \hspace{2cm}
\end{multline}
We also have $\theta=\rho-1-\overline{\eta}$ leading to $\theta=1/4+\mathcal{O}(\epsilon^2)$. 
Note that $\overline{\eta}$ depends on $\epsilon$ while $\eta$ sticks to its mean-field value $\eta=0$ also for $\epsilon>0$ as usual for long-range interactions \cite{Fisher_1972}. At finite temperature, it has been suggested that there is a correspondence between the critical exponents of short-range models in dimension $D$ and those of long-range models for some appropriate $\rho=\rho(D)$ \cite{Kotliar_1983,katzgraber2009study,Moore_2010,Martin-Mayor_2012,Angelini_2014_relations}. The arguments are qualitative and the correspondence is approximate, nonetheless, it sometimes holds at first order in the $\epsilon$ expansion \cite{Kotliar_1983,Martin-Mayor_2012}. In the present context, the arguments lead to the following formulas: $\nu_{\text{SR}}=\nu_{\text{LR}}/D$,  $\theta_{\text{SR}}=D\,\theta_{\text{LR}}$, $\omega_{\text{SR}}=D\,\omega_{\text{LR}}$ with $\rho=D+2-\eta(D)$. An explicit computation shows that \eqref{eq:short_range_exponents}  and \eqref{eq:long_range_exponents} follow the correspondence in the first order in $\epsilon$.

\section{Clusters and avalanches}
\label{sec:cluseava}
In order both to clarify some physical points that we mentioned earlier on but also to connect the present framework to our previous work \cite{angelini2022unexpected,angelini2020loop}, we discuss the relation betweeen clusters as introduced in the previous section and avalanches.
Avalanches are a natural and intuitive concept that is often used in the literature on disordered systems at zero temperature see e.g. \cite{monthus2011random,tarjus2013avalanches}.Indeed, an intuitive way of studying zero-temperature correlations between two spins $\sigma_i$ and $\sigma_j$ on the lattice is to consider the difference between the ground state of the system and the new ground state in which spin $\sigma_i$ is constrained in the direction $-h_i$. 
If in the new ground state, $\sigma_j$ is flipped, then we may say that the two spins are correlated and define a response function $R_{ji}$ equal to one in this case and zero otherwise, note that this is the definition used in \cite{angelini2022unexpected,angelini2020loop}. We call the avalanche of spin $i$ the set of spins that flip when $i$ is flipped, i.e. all the spins such that $R_{ji}=1$. 
A problem with defining correlation in terms of $R_{ji}$ is that we can find instances of the disorder in which spin $\sigma_j$ flips if we flip $\sigma_i$ but not vice-versa, i.e. it may happen that $R_{ij} \neq R_{ji}$, meaning that responses are not symmetric.

In order to have a symmetric definition of the correlation function, we recall the effective energy function $E_{ij}(\sigma_i,\sigma_j)$ defined in Sec. (\ref{sec:model}). This function specifies, for a given realization of the disorder, the ground state configuration and the three excited configurations with the corresponding excitation energies.
Let us define a quantity, $C_{ij}$, that is equal to one if the configuration with the lowest excitation energy among the three possible ones is the one where both spins are flipped.
Obviously $C_{ij}=1$ implies $R_{ij}=R_{ji}=1$ and $\Delta E_i=\Delta E_j=\Delta E$, where $\Delta E$ is the excitation energy of the first excitation. This means that, if $C_{ij}=1$, spins $i$ and $j$ are in the same cluster of spins with local fields $|h_i|=|h_j|=\Delta E/2$. 

We now want to show that the opposite is true, i.e. that if $|h_i|=|h_j|$ then $C_{ij}=1$. To be precise, this is true with probability one over the disorder realization. By definition, the modulus of the local field $|h_i|$($|h_j|$) is equal to (half) the energy difference $\Delta E_i$ ($\Delta E_j$) of the lowest configuration $c_i$ ($c_j$) in which spin $i$ ($j$) is flipped with respect to the ground state.  Also by definition, if $C_{ij}=1$, these configurations are the same, $c_i=c_j$, while if $C_{ij}=0$ they are distinct, $c_i \neq c_j$. Since either the $J_{ij}$'s or the $H_i$'s obey a continuous distribution, the three excitation energies have a continuous distribution as well, meaning that the probability that a given excitation energy lies in a narrow interval $(\Delta E-dE/2,\Delta E+dE/2)$ around $\Delta E$ is $\mathcal{O}(dE)$. Therefore, if  $c_i \neq c_j$ the probability that both the corresponding excitation energies $\Delta E_i$ and $\Delta E_j$ lie in the same narrow interval around $\Delta E$ is $\mathcal{O}(dE^2)$ while if $C_{ij}=1$ the probability that the excitation energy of the single configuration $c_i=c_j$ lies in a narrow interval around $\Delta E$ is $\mathcal{O} (dE)$.
It follows that if $\Delta E_i$ and $\Delta E_i$ lie in the same narrow interval around $\Delta E$, with probability one they actually coincide, i.e. $C_{ij}=1$.
The connection with excited configurations can be generalized: given a cluster of size $s$ with excitation energy $\Delta E$, the lowest of the $2^s-1$ excitations of the restricted system of the $s$ spins in the cluster is the one in which all the $s$ spins are flipped with respect to the ground state. 

Concerning clusters and avalanches, we recall that we may have that $R_{ij}\neq R_{ji}$ therefore the avalanche of a given spin contains the cluster of the spin but may be larger. The following statements can be shown to hold: i) all spins in a cluster have the same avalanche, i.e. for each cluster there is one and only one avalanche. ii) the excitation energy of any spin $j$ in the avalanche of a given cluster must be smaller or at most equal to the excitation energy of the cluster $\Delta E_j \leq \Delta E$.  This last statement implies notably that clusters and avalanches coincide if $\Delta E =0$, i.e. in the case of soft clusters, most relevant to this work.

Finally, because of the analogy with percolation, it is important to remark that it is possible to find clusters that are \quotationmarks{disconnected} in the sense of percolation. By this, we mean that we can find clusters such that there is at least one couple of spins in the set not connected by a linear path made of spins belonging to the set.
This disturbing feature is not present if we consider avalanches, because a spin may flip only if at least one of its neighbors flips. On the other hand, since clusters coincide with avalanches for $\Delta E=0$, it follows that soft clusters are also connected in the sense of percolation.

\matmethods{

\subsection{The $M$-layer construction}
In this section, we recall the diagrammatic rules derived in ref. \cite{Altieri_2017} to compute a generic $q$-points observable, averaged over the possible rewirings, inside the $1/M$ expansion, referring to the original paper \cite{Altieri_2017} and to the pedagogical applications in refs. \cite{Angelini_2024Ising, angelini2025bethe} for their complete derivation and additional details.
The procedure is composed of the following steps:
\begin{enumerate}
\item Identify the relevant topological diagrams:
 In the limit of large $M$, in a quenched realization of the random rewirings, if the $q$ sites are connected, they will be connected by a sequence of adjacent edges without topological loops: the leading order contribution to the $1/M$ expansion for the chosen observable will be given by this type of diagrams. To be concrete, the leading order contributions for the two- and three-point susceptibilities come respectively from the left diagram in Fig. \ref{fig:2pointSGH} and \ref{fig:3pointSGH}.
If then we want to compute the next-to-leading order, we need to include also diagrams that have an additional topological loop, that will bring an additional factor $1/M$. The next-to-leading order contributions for the two- and three-point susceptibilities come respectively from the right diagram in Fig. \ref{fig:2pointSGH} and \ref{fig:3pointSGH}.
Thus one needs to associate to each diagram a factor that is a power of $1/M$, describing the probability that
a topological diagram of that kind is obtained in the rewiring procedure.
\item Compute factors for each diagram:
In addition to the factor in powers of $1/M$, which we will call $W(\mathcal{G})$, one needs to associate to each diagram $\mathcal{G}$ a symmetry factor $S(\mathcal{G})$, that takes the same form as the one introduced for Feynman diagrams in field theory \cite{zinn2021quantum}, and a factor $\mathcal{N}(\mathcal{G})$ that counts the number of possible realizations with the same form of the chosen topological diagram on the original lattice.
\item Compute the line-connected observable on the chosen diagram:
For any chosen diagram, one needs to compute the observable on a Bethe lattice in which the topological structure of that
diagram has been manually injected. If one loop is present, to avoid multiple counting, one then needs to compute the line-connected observable, that is the value of the observable on the given diagram from which we subtract all the contributions obtained computing the observable on diagrams where a single line composing the loop is removed (we refer to ref. \cite{Altieri_2017} for the case in which more than one loop is present).
\item Sum of the contributions:
In the end, we sum the contributions to the chosen observable coming from the different chosen diagrams, multiplying the value
of the line-connected observable on a given diagram by the factors associated with that diagram and summing over the
positions of internal vertices and the lengths of the internal lines.
\end{enumerate}

\subsection{Critical Exponents below $\DU=8$}

We have computed the susceptibilities up to one loop order in powers of $1/M$, following the prescriptions of the precedent section, using for $\chi_2$ and $\chi_2^{dis}$ the diagrams in Fig. \ref{fig:2pointSGH} and for $\chi_3$ the diagrams in Fig. \ref{fig:3pointSGH}. Diagrams involving tadpoles can be shown to be irrelevant with the help of the condition given by \eqref{cond1}, see section C.1. in the SI Appendix.
\begin{figure}[H]
    \centering
    \includegraphics[scale=0.25]{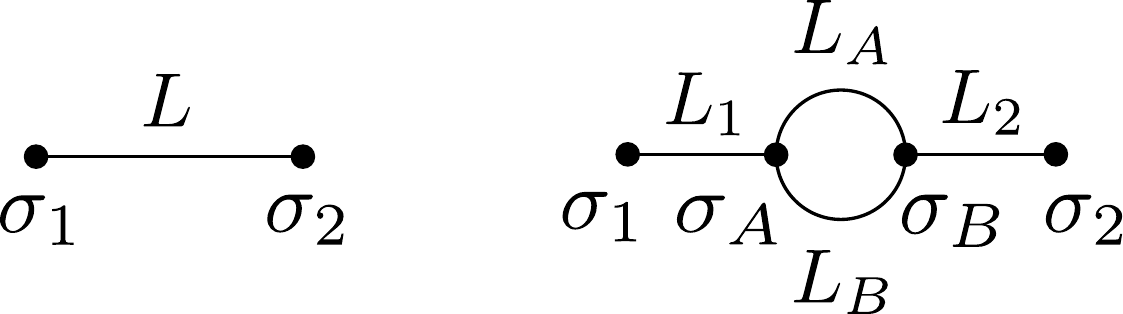}
    \caption{Diagrams considered for the computation of the observables $\chi_2$ and $\chi_2^{dis}$ up to one-loop order. The diagram on the left gives the leading $\mathcal{O}(1/M)$ contribution, and the diagram on the right gives the first $\mathcal{O}(1/M^2)$ correction.}
    \label{fig:2pointSGH}
\end{figure}
\begin{figure}[H]
    \centering
    \includegraphics[scale=0.25]{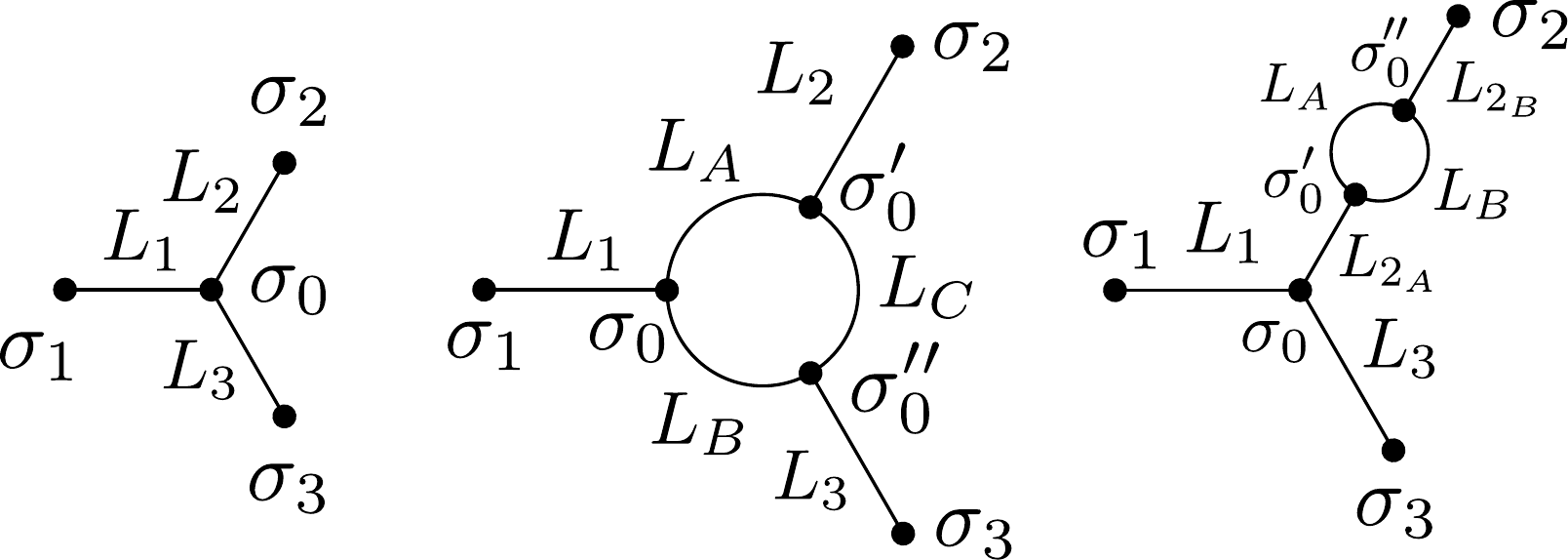}
    \caption{Diagrams considered for the computation of the observable $\chi_3$ up to one-loop order. From left to right the diagrams give contributions of order $\mathcal{O}(1/M^2)$, $\mathcal{O}(1/M^3)$, $\mathcal{O}(1/M^3)$.}
    \label{fig:3pointSGH}
\end{figure}
The results, expressed in terms of $m\equiv \xi^{-1}$ 
, are:
\begin{equation}
    \chi_2\propto \,m^{-2}
    \left(       1    +  
       \frac{1}{(4\pi)^{D/2}}\frac{u}{2}\,I_4           
    +\mathcal{O}\left(u^2\right)\right)      \,,
\label{eq:chi2}
\end{equation}
\begin{equation}
    \chi_3 \propto m^{-6} 
    \left(       1    +  \frac{u}{(4\pi)^{D/2}}\left( \frac{3}{2}I_4 -I_3        \right)  +\mathcal{O}\left(u^2\right)\right)    \,,
\label{eq:chi3}
\end{equation}
\begin{equation}
    \chi_2^{dis} \propto m^{-4}
    \left(       1    +  \frac{u}{(4\pi)^{D/2}}\left( I_4-I_2       \right)  +\mathcal{O}\left(u^2\right) \right)    \,,
\label{eq:chi2dis}
\end{equation}
where $u \equiv g\, m^{D-8}$ and $g$ is a $\mathcal{O}(1/M)$ quantity that depends on the microscopic details of the model, see the SI Appendix. As we will now see its actual value is irrelevant for the computation of the critical exponents leading to universal behavior.
\footnote{The expansions in Eqs. (\ref{eq:chi2},\ref{eq:chi3},\ref{eq:chi2dis}) depend on a single microscopic $\mathcal{O}(1/M)$ parameter $u$. To be fair in the parenthesis there are other $\mathcal{O}(1/M)$ corrections that depend on additional model-dependent parameters. These corrections, however, can be safely neglected at the critical point $m \approx 0$ because they are sub-leading in $m$ with respect to $u \propto m^{D-8}$.}.
We also have:
\begin{equation}
    I_2\equiv    \int^\infty_{0} \int^\infty_{0}\,dL_a\,dL_b\frac{L_a^2L_b\,e^{-(L_a+L_b)}}{(L_a+L_b)^{D/2+1}}\, =
     \frac{1}{12}\Gamma\left(\frac{\epsilon}{2}\right)\,,
\end{equation}
\begin{multline}
    I_3  \equiv  \int^\infty_{0}\int^\infty_{0}\int^\infty_{0} dL_a\,dL_b\,dL_c\frac{L_aL_b+L_cL_b+L_aL_c}{(L_a+L_b+L_c)^{D/2+1}}\, \times \\ 
      e^{-(L_a+L_b+L_c)}\,  = \frac{1}{8}\Gamma\left(\frac{\epsilon}{2}\right)\,,
\end{multline}
\begin{equation}\label{eq:I4}
     I_4 \equiv  \int^\infty_{0}\int^\infty_{0} dL_a\,dL_b\frac{L_a^2L_b^2e^{-(L_a+L_b)  }}{(L_a+L_b)^{D/2+2}}  
    = \frac{1}{30}\Gamma\left(\frac{\epsilon}{2}\right)\,,
\end{equation}
with $\epsilon \equiv 8-D$.
Inserting Eqs. (\ref{eq:chi2},\ref{eq:chi3},\ref{eq:chi2dis}) into the definition of the renormalized coupling constant, \eqref{def:lambda}, we obtain at second order in $u$:
\begin{equation}
    \lambda=u  - \frac{4}{15}\frac{u^2}{(4\pi)^{D/2}} \Gamma\left(4-\frac{D}{2}\right) +\mathcal{O}(u^3)\,.
\label{eq:lambdau}
\end{equation}
According to the scaling laws, $\lambda$ has a finite value $\lambda_c$ at the critical point for $D<8$, but we cannot easily read that value from the above series because $u$ instead diverges at the critical point $m=0$, according to its definition  $u \equiv g\, m^{D-8}$. Therefore it is convenient to transform the series in powers of $u$ into series in powers of $\lambda$. To determine the critical value of $\lambda$ we follow ref. \cite{Parisi1988}, Chap. 8, and introduce the $\beta$ function as:
\begin{equation}
    \beta \equiv m^2 \frac{\partial}{\partial m^2}\Bigg|_{g \text{ fixed}}\lambda=\frac{1}{2}(D-8)u\frac{\partial}{\partial u}\Bigg|_{m^2 \text{ fixed}}\lambda\,.
    \label{betafunction}
\end{equation}
From \eqref{eq:lambdau} and \eqref{betafunction} we obtain the $\beta$ function in powers of $u$, and then, computing $u$ as a function of $\lambda$, inverting \eqref{eq:lambdau}:
\begin{equation}
    u=\lambda + \frac{4}{15}\frac{\lambda^2}{(4\pi)^{D/2}} \Gamma\left(4-\frac{D}{2}\right) +\mathcal{O}(\lambda^3)\,,
\label{eq:ulambda}
\end{equation}
we get to second order in $\lambda$:
\begin{equation}
    \beta(\lambda)=-\frac{\epsilon \, \lambda}{2}+ \frac{2 \, \epsilon}{15 (4\pi)^{D/2}} \Gamma\left(\frac{\epsilon}{2}\right) \lambda^2+\mathcal{O}(\lambda^3)\,.
    \label{eq:betalambda}
\end{equation}
Since, at the critical point, $\lambda$ converges to some $\lambda_c$ we must have $\beta(\lambda_c)=0$. More precisely we expect that close to the critical point, where $m$ is small, $\lambda = \lambda_c+ c_1 m^{- \omega}$ with a universal negative exponent $\omega$ that controls the leading finite-size corrections to scaling. This implies:
\begin{equation}
\beta(\lambda_c)=0 \ ,\ \ \  \omega=-2\, \beta'(\lambda_c)\,.
\end{equation}
From \eqref{eq:betalambda} the following scenario emerges for the zeroes of the beta function: for $\epsilon \leq 0$ ($D \geq 8$)  only the solution 
$\lambda_c=0$ exists, meaning that $\lambda$ tends to zero at the critical point with $\omega=\epsilon+\mathcal{O}(\epsilon^2)$ (in agreement with the discussion after \eqref{nhighd}), while for $\epsilon>0$ a new solution $\lambda_c > 0$ appears while the solution $\lambda_c=0$ becomes unstable as $\omega=\epsilon+\mathcal{O}(\epsilon^2)$ would be positive, thus $\lambda_c=0$ is un-physical. Expanding at small $\epsilon$,  we obtain by means of $\lim_{\epsilon \rightarrow 0} \epsilon \, \Gamma(\epsilon)=1  $ 
\begin{equation}
  \lambda_c = \frac{15}{8}(4\pi)^{4}  \epsilon + \mathcal{O}(\epsilon^2)
\end{equation}
and $\omega=-\epsilon+\mathcal{O}(\epsilon^2)$.
The critical exponents $\eta$ and $\overline{\eta}$ can be evaluated considering the following effective exponents functions:
\begin{equation}
    Q(\lambda) \equiv  \frac{\partial \ln \chi_2}{\partial \ln m^2}\Bigg|_{g \ \text{fixed}} \,  , \ \ \   Q^{dis}(\lambda) \equiv  \frac{\partial \ln \chi_2^{dis}}{\partial \ln m^2}\Bigg|_{g \ \text{fixed}} \, .
\end{equation}
They can be obtained from Eqs. (\ref{eq:chi2},\ref{eq:chi2dis}) and expressed in powers of $\lambda$ from \eqref{eq:ulambda}. 
The scaling laws $\chi_2 \propto m^{\eta-2}$, $\chi_2^{dis} \propto m^{\overline{\eta}-4 }$ for $m \rightarrow 0$ imply that $\eta$ and $\overline{\eta}$ are related to the effective exponents functions  evaluated at the critical point $m=0$ (i.e. $\lambda =\lambda_c$): 
\begin{equation}
   Q(\lambda_c)= -1 + \frac{\eta}{2} \  \ ,\ \   Q^{dis}(\lambda_c)= -2 + \frac{\overline{\eta}}{2} \ .
\end{equation}
The exponent $\nu$ can be computed considering $\partial  t/\partial m^2$. Since $m \propto t^{\nu}$ it follows that $\partial  t/\partial m^2 \propto m^{1/\nu-2}$ for $m\rightarrow 0$ or equivalently $\lambda \rightarrow \lambda_c$. The $M$-layer computation leads to:
\begin{equation}
 \frac{\partial  t}{\partial m^2} \propto1-\frac{1}{(4\pi)^{D/2}}\frac{u}{2}\left(\frac{D-16}{60}\right)\Gamma\left(\frac{\epsilon}{2} \right)\,+\mathcal{O}(u^2).
\end{equation}
to be combined with the following expressions:
\begin{equation}
    c_2(\lambda) \equiv  \frac{\partial \ln (\partial t/\partial m^2)}{\partial \ln m^2}\Bigg|_{g \ \text{fixed}} \, , \ \ \   c_2(\lambda_c)=\frac{1}{2\, \nu} -1\ .
\end{equation}

}

\showmatmethods{} 

\section{Conclusions}

Until now no perturbative stable RG fixed point (FP) was found for the spin glass model in a field below the upper critical dimension $\DUFC=6$. Several authors then look at this absence as an indication of the disappearance of the SG phase below the upper-critical dimension.
In this work, we follow a perturbative RG approach through a loop-expansion around the Bethe lattice solution of the model, and we are able to find a perturbative FP, different from the MF one, that is stable below $\DU=8$. This FP is a $T=0$ one and thus has different properties compared to standard $T\neq 0$ FPs. In particular, the associated independent critical exponents are three and not just two. We thus computed the exponents $\nu$, $\eta$ and $\overline{\eta}$ inside an $\epsilon$-expansion around $D=8$.
Our computations are done directly at $T=0$. This would be impossible using standard field-theoretical methods both because the MF FC SG model around which one expands has no transition in field exactly at $T=0$, and because the Lagrangian is not well defined at $T=0$.
The $M$-layer expansion that we used instead is well defined even at $T=0$, and moreover the expansion is performed around the Bethe lattice solution that has a $T=0$ transition at a critical value of the field.
One could wonder if the results that we obtained are valid also for $T>0$. Following common folklore, if the exponent $\theta$ is positive, the temperature should be an irrelevant parameter and the critical line at $T>0$ should be controlled by the $T=0$ FP. The exponents measured at finite temperature should thus be the ones associated with the $T=0$ fixed point, as first conjectured for the RFIM \cite{Bray_1985} and then numerically verified.

We thus believe that now it is crucial to perform more precise numerical simulations, both on SR and LR models, to numerically check our estimates for the critical exponents.
In order to compute numerically the functions $P^{con}(x_1,x_2;u,u',J)$ and $P^{dis}(x_1,x_2;u,u')$ we should compute the triplet $(u,u',J)$ between spins at positions $x_1$ and $x_2$ on a given realization of the disorder and then average over many realizations.
To do this we should compute the ground state configuration of the whole system and the three excited states obtained forcing the two spins in the three possible configurations other than the ground state. Inverting equation $E_{ij}(\sigma_i,\sigma_j)= u\, \sigma_{i}+ u'\,\sigma_{j}+J\, \sigma_i\, \sigma_j$ one obtains the triplet $(u,u',J)$. This procedure is computationally demanding, alternatively one could consider standard finite-temperature simulations.
Indeed, while a detailed analysis of the implications of the present work for finite-temperature observables is left for future work, we expect that the exponents $\nu$ and $\eta$, measured at $T\neq0$ from the correlation length and from the two-point susceptibilities correspond to the ones we computed at $T=0$. This is what happens in the RFIM.  Besides we remark that, according to our results, $\eta$ is negative and $\nu$ is larger than $1/2$ below the upper critical dimension, in qualitative agreement with the finite-temperature numerical estimates  $\eta \approx -0.3$ and $\nu \approx 1.46$ in four dimensions ($\epsilon=4$) \cite{banos2012thermodynamic}.
Numerical data in dimension six could not be used to obtain estimates, however, they are compatible with our results inside the numerical error \cite{aguilar2024evidence}. 
In order to extract the other exponents, we note that Eqs. (\ref{eq:X2scalb}) should remain valid for $T>0$ as well. Thus, a finite-temperature cubic connected function such as those measured in \cite{fernandez2022numerical} should obey the scaling laws of $\chi_3$ and therefore on could extract $\theta$ from the ratio $\chi_3^2/\chi_2^3 \propto \xi^{D-\theta}$ or similar.
Apparently there are no finite temperature correlations that tend to the disconnected zero-temperature correlations we defined in the limit $T=0$ (standard disconnected correlation functions as defined at $T>0$ go to zero in the $T\to 0$ limit); however, one can extract the exponent $\overline{\eta}$ from the numerics determining $\eta$ and $\theta$, as suggested above.
A detailed analysis of the implications of the present work for finite-temperature observables is left for future work. 

Finally, let us mention the fact that for $D>6$, also the Gaussian FP associated with the FC transition is stable (although it has a finite basin of attraction). Thus, if temperature is an irrelevant parameter in the vicinity of $T=0$, for $6<D<8$ there are two stable FPs that could attract the RG flow. It could be possible that the RG flow on the critical line goes to the FC FP or to the $T=0$ FP depending on the value of the field. Numerical investigation is crucial to understand this point.

\acknow{We thank M. A. Moore and M. Aguilar-Janita, V. Martin-Mayor, J. Moreno-Gordo, and J. J. Ruiz-Lorenzo for interesting discussions and
for sharing their numerical data. This project has been supported by funding
from the 2021 first FIS (Fondo Italiano per la Scienza) funding scheme (FIS783
- SMaC - Statistical Mechanics and Complexity) from
Italian MUR (Ministry of University and Research).}

\showacknow{} 

\bibsplit[4]

\bibliography{biblio}

\appendix


\section*{Supplementary material}

# Supporting Information for "Critical exponents of the spin glass transition in a field at zero temperature"

## Contents

In this Supporting Information (SI), we detail the computations that lead to the expressions for the susceptibilities, which serve as the starting point for calculating the critical exponents discussed in the main text. Section 1 focuses on the mean-field analysis of the spin glass in a field at zero temperature. Next, in Section 2 we apply the $M$-layer construction to the model of interest. Finally, in Section 3, we present numerical tests that validate our analytical non-mean-field results for the observables on the $M$-layer lattice.

## 1. Mean-field behavior on the Bethe lattice

The first step in studying the critical behavior of a model in finite dimensions is to analyze the corresponding mean-field approximation. For the spin glass model in a field we know that a zero-temperature transition occurs in the Bethe lattice for a finite value of the external field (32). In this SI, we restrict our analysis to the zero-temperature case. Throughout the following, it will be assumed that all computations are performed in this regime. Here we compute a useful approximation of the joint probability distribution of the two effective fields and the effective coupling between two generic spins at a large distance in the Bethe lattice. We can perform this computation by exploiting the locally tree-like structure of this topology. Using this approximated distribution we can estimate the scaling functions cited in the main text and the mean-field expressions of the observables of interest.

**A. Semi-analytical *Ansatz* on the Bethe lattice.** We now consider a Bethe lattice with connectivity $2D$, where $D$ is the dimension of the associated $M$-layer graph, as defined in the main text. Considering the Hamiltonian given in Eq. (1) of the main text and in particular the case of Gaussian couplings and fixed positive magnetic field, $H$, on a Bethe lattice one can associate to any edge between nodes $i$ and $j$, a cavity field $u_{j \to i}$. The cavity fields are defined such that, knowing them, one can extract the marginal probability distribution of spin $\sigma_i$ simply as:

$$\mu_i(\sigma_i) = \frac{1}{\mathcal{N}} e^{-\beta\left(H + \sum_{j \in \partial i} u_{j \to i}\right)\sigma_i}, \qquad [44]$$

where $\mathcal{N}$ ensures the normalization of the marginal probability and we indicate with $\partial i$ the set of nearest neighbors of node $i$. In the Bethe lattice at zero temperature the distribution of the cavity fields $P_B(u_i)$ over the disorder, which we call "Bethe cavity distribution", obeys the following implicit equation (40, 58)

$$P_B(u) = \mathbb{E}_J \int \prod_{i=1}^{2D-1} P_B(u_i) \, du_i \, \delta\left(u - \text{sign}\left(J\left(H + \sum_{i=1}^{2D-1} u_i\right)\right) \min\left(|J|, |H + \sum_{i=1}^{2D-1} u_i|\right)\right). \qquad [45]$$

As already stated in the main text we expect that, due to universality, the same results obtained in this SI hold for the case of bimodal couplings and Gaussian external fields $H_i$: in the following, we will consider fixed $H$ and Gaussian couplings.

To begin, we focus on computing the correlation between two spins, $\sigma_1$ and $\sigma_2$. These spins are connected by a unique sequence of adjacent edges, referred to as a *path* or *chain*. This path can be effectively characterized by three parameters: two effective fields, $u_1$ and $u_2$, acting on $\sigma_1$ and $\sigma_2$, respectively, and an effective coupling $J$.

A convenient way to describe the path connecting $\sigma_1$ and $\sigma_2$ is to consider only the sites and edges directly linking them while imagining that each internal site in the chain is connected to the two neighboring sites along the chain and to $2D - 2$ infinite trees from outside the chain. This construction preserves the original connectivity of the Bethe lattice, $2D$, on the internal spins. Notably, this approach implies that the external spins have connectivity equal to 1, a feature that proves useful in the following computation, as discussed few lines below.

Each of these infinite trees contributes a cavity bias $u$ distributed according to $P_B(u)$. Summing over all internal spin configurations, the entire chain can then be effectively described by the three parameters mentioned above.

As done in ref. (40), one can introduce the distribution of the triplet $(u_1, u_2, J)$ for two sites at distance $L$, that we call "semi-analytical *Ansatz*"

$$P_L(u_1, u_2, J) \equiv \delta(J) \, P_B(u_1) \, P_B(u_2) - 2 \, a \, L \, \lambda^L \, \delta(J) \, g(u_1) \, g(u_2) + a \, L^2 \, \lambda^L \, \rho \, e^{-\rho|J|L} \, g(u_1) \, g(u_2), \qquad [46]$$

where $a$ and $\rho$ are parameters that can be computed, while $\lambda$ and $g(u)$ are, respectively, the largest eigenvalue and the associated eigenfunction of the following integral equation (32):

$$\lambda\, g(u) = \mathbb{E}_J \int \prod_{i=1}^{2D-2} P_B(u_i) du_i\, g(u') du'\, \mathbb{I}\!\left[\Big|H + u' + \sum_{i=1}^{2D-2} u_i\Big| < |J|\right] \delta\!\left(u - \mathrm{sign}(J)|H + u' + \sum_{i=1}^{2D-2} u_i|\right), \qquad [47]$$

being $\mathbb{I}[\,\cdot\,]$ the indicator function. The function $g(u)$ satisfies $g(u) = g(-u)$, $\int du\, g(u) = 1$ and $g(u) \geq 0$. In this context $\lambda$ is the control parameter, at the critical point $H = H_c$ it assumes the critical value $\lambda_c = 1/(2D-1)$. Notice also that three terms contribute to the distribution in Eq. (46): the first is the trivial case in which the effective coupling is zero and the cavity fields are distributed according to $P_B(u)$, that is the only term that survives in the limit $L \to \infty$; the second and third contributions give the corrections to it. In particular, the third term is an exponential distribution that is expected for large lengths, while the second term represents the finite weight for $J = 0$.

This distribution is valid in the large-length regime $L \gg 1$. The expression of $P_L(u_1, u_2, J)$ is indeed obtained writing a generic form up to order $L\lambda^L$

$$P_L(u_1, u_2, J) = \delta(J)\left[P_B(u_1)P_B(u_2) - b\, L\lambda^L g(u_1)g(u_2) - c_1 L\lambda^L g'(u_1)g'(u_2) - c_2 L\lambda^L g''(u_1)g''(u_2)\right] + $$
$$+ a L^2 \lambda^L \rho\, e^{-\rho|J|L} g(u_1) g(u_2)\,, \qquad [48]$$

and then, joining two lines and imposing "self-consistency", the free parameters can be fixed (40), leading to the expression given in Eq. (46). Here, "self-consistency" refers to the requirement that the distribution obtained by joining two lines of lengths $L_1$ and $L_2$ must have the same form as the distributions of the original lines, but with the length parameter equal to $L_1 + L_2$.

In order to join two chains with external spins $\sigma_1$ and $\sigma_2$, we should marginalize over the spin $\sigma$ connecting them. At $T=0$ the only configuration that contributes is the one minimizing the energy, thus it depends on the field $h$ acting on $\sigma$ and the two couplings $J_1$ and $J_2$ that link $\sigma$ with the neighboring spins, respectively $\sigma_1$ and $\sigma_2$. As an output of this operation, we have two effective fields acting on the spins $\sigma_1$ and $\sigma_2$ and an effective coupling $J_{12}$ between them:

$$-u_1\,\sigma_1 - u_2\,\sigma_2 - \sigma_1\, J_{12}\,\sigma_2 \equiv \min_\sigma(-\sigma_1\, J_1\, \sigma - \sigma_2\, J_2\, \sigma - h\,\sigma)\,. \qquad [49]$$

In Tab. S1 we report the rules for this marginalization, where, $h$, $J_1$, $J_2$ and $u_1$, $u_2$, $J_{12}$ are the input and output parameters respectively:

**Table S1. Rules for evolution of a straight line in the case $|J_2| \geq |J_1|$. The complementary case $|J_1| \geq |J_2|$ can be obtained simply exchanging index** 1 **with index** 2, *i.e.* $u_1$ **with** $u_2$ **and** $J_1$ **with** $J_2$.

|  | $u_1$ | $u_2$ | $J_{12}$ |
|---|---|---|---|
| $|h| > |J_2| + |J_1|$ | $\mathrm{sign}(h)J_1$ | $\mathrm{sign}(h)J_2$ | 0 |
| $|J_2| - |J_1| < |h| \wedge$ $|h| < |J_2| + |J_1|$ | $\mathrm{sign}(J_1)h_-(J_1, J_2)$ | $\mathrm{sign}(J_2)h_+(J_1, J_2)$ | $\mathrm{sign}(J_1 J_2)\widetilde{J}$ |
| $|h| < |J_2| - |J_1|$ | 0 | $\mathrm{sign}(J_2)h$ | $\mathrm{sign}(J_2)J_1$ |

with $\widetilde{J} = \frac{|J_1| + |J_2| - |h|}{2}$ and $h_\pm(J_1, J_2) = \mathrm{sign}(h)\frac{|h| \pm (|J_2| - |J_1|)}{2}$.

We remark that the *Ansatz* distribution on the effective triplet characterizes the effective parameters acting on two spins (uniquely) connected by $L-1$ spins with connectivity $2D$, while the two external spins themselves have connectivity equal to 1. This choice ensures that lines can be joined by adding an arbitrary number of cavity fields on the joining site, allowing to fix its connectivity to the chosen value $2D$. For instance, when connecting $q$ lines on a central spin $\sigma_0$ to form a $q$-degree vertex, to retrieve connectivity equal to $2D$ for the spin $\sigma_0$, we only have to add the contributions of $2D - q$ cavity fields $u_i$ to it, with $i = 1, \dots, 2D - q$.

As a last comment we want to connect the expression for the distribution of the triplet in Eq. (5) of the main text with the expression on the Bethe lattice, reported here, in Eq. (46). The former describes the effective triplet on the (finite-dimensional) $M$-layer lattice, thus additional information regarding the spatial positions is needed, while on the Bethe lattice, there is no notion of Euclidean space. Moreover, when working on tree-like graphs it is useful to consider cavity distributions, as $P_B(u)$ for instance, that differs from the distribution $P_1(h)$ in the main text, where $h$ is the "total" local field acting on a single site, not a cavity quantity as the bias $u$. In Sec. A, we will show how to use this distribution to compute observables on more complicated topologies, including those with loop structures.

In order to obtain the triplet distribution on the Bethe lattice we can attach to the external spins $2D - 1$ fields, distributed as $P_B(u)$, and the external field $H$. From Eq. (46) we obtain:

$$Q_L(u_1, u_2, J) - \delta(J) P_1(u_1) P_1(u_2) = \widehat{P}_1(u_1) \widehat{P}_1(u_2) a\left(-2\delta(J) L\lambda^L + \rho L^2 \lambda^L e^{-\rho|J|L}\right), \qquad [50]$$

where $P_1(u)$ is the "total" local field distribution on the Bethe lattice and $\widehat{P}_1(u_1)$ is defined from:

$$\widehat{P}_n(h) \equiv \int \prod_{i=1}^{n} g(u_i) du_i \prod_{j=n+1}^{2D} P_B(u_j) du_j\, \delta\!\left(h - (H + \sum_{i=1}^{n} u_i + \sum_{j=n+1}^{2D} u_j)\right) \qquad [51]$$

in the specific case $n = 1$. We notice that $\widehat{P}_2(h)$ satisfies the following relationship

$$\frac{4\, a\, \widehat{P}_2(0)}{\rho} = 1\,, \qquad [52]$$

that follows from the self-consistency of the *Ansatz* distribution (40). Notice that, according to the previous definition, we can identify $\widehat{P}_0(h)$ with $P_1(h)$ on the Bethe lattice.

**B. Scaling functions on the $M$-layer lattice.** Before moving to the actual computations let us compute the distribution of the triplet on the $M$-layer graph at leading order (corresponding to the Bethe approximation), averaged over the random rewirings and the disorder. In order to do so, following the prescriptions of the $M$-layer construction (29), we have to include the number of "Non-Backtracking Paths" (NBP), $\mathcal{N}_L(x,x')$, between two points $x$ and $x'$ and sum over the length of the path, $L$, with the corresponding weight $W = 1/M$

$$P_2(x,x';u,u',J) - \delta(J)P_1(u)P_1(u') = \frac{1}{M}\sum_{L=1}^{\infty} \mathcal{N}_L(x,x') \left( Q_L(u,u',J) - \delta(J)P_1(u)P_1(u') \right). \quad [53]$$

The number of NBP in a $D$-dimensional hypercube is:

$$\mathcal{N}_L(x,x') = \left(\frac{2D}{2D-1}\right)\frac{(2D-1)^L}{(4\pi L/c)^{D/2}} e^{-c\frac{(x-x')^2}{4L}} \quad [54]$$

where $c = \frac{2D-2}{a_l^2}$ and $a_l$ is the lattice spacing, in the remainder of this section we will set $c=1$ for simplicity with an appropriate rescaling of the length. Going to Fourier space we have

$$FT[\mathcal{N}_L(x,x')] \propto \delta(k+k')(2D-1)^L e^{-k^2 L}, \quad [55]$$

from which we compute the Fourier transform of the distribution of the triplet:

$$FT[P_2(x,x';u,u',J) - \delta(J)P_1(u)P_1(u')] \propto \delta(k+k')\frac{1}{M}\sum_{L=1}^{\infty}(2D-1)^L e^{-k^2 L} \widehat{P}_1(u)\widehat{P}_1(u')\left(-2\delta(J)L\lambda^L + \rho L^2 \lambda^L e^{-\rho|J|L}\right) =$$

$$\delta(k+k')\frac{1}{M}\sum_{L=1}^{\infty} e^{-k^2 L - \tau L} \widehat{P}_1(u)\widehat{P}_1(u')\left(-2\delta(J)L + \rho L^2 e^{-\rho|J|L}\right) =$$

$$\delta(k+k')\frac{1}{M}\widehat{P}_1(u)\widehat{P}_1(u')\left(-\delta(J)\frac{2}{(k^2+\tau)^2} + \rho\frac{2}{(k^2+\tau+\rho|J|)^3}\right), \quad [56]$$

where we defined the distance from the mean-field critical point $\tau \equiv -\ln(\lambda(2D-1))$ and we assumed that we are near the critical point $\tau \approx 0^+$ and at large distances $k \approx 0$. Measuring $J$ in units of $\rho$ we obtain Eq. (6) of the main text. Note indeed that, at leading order in $1/M$, $t$ (defined in the main text) is proportional to $\tau$. Notice that at the mean-field level we can identify the correlation lengths, which are diverging at the critical point on the $M$-layer lattice, as $\xi \propto \tau^{-1/2}$ and $\xi_{con} \propto (\tau + \rho|J|)^{-1/2}$, while if we take into account loop corrections these relations are more complicated, as shown in the main text and in Sec. C of this SI.

In real space we obtain

$$P^{dis}(x_1,x_2;u_1,u_2) = \frac{1}{M}\sum_{L=1}^{\infty} \mathcal{N}_L(x_1,x_2) P_L^{J=0} \widehat{P}_1(u)\widehat{P}_1(u') \propto \int_1^{\infty} dL\, L^{1-D/2} e^{-\tau L - \frac{(x_1-x_2)^2}{4L}} \widehat{P}_1(u)\widehat{P}_1(u') \quad [57]$$

where we replaced the sum over the length with the corresponding integral. Following the scaling expression in Eq. (7) of the main text we can find the scaling expression near the critical point and for large $r = |x_1 - x_2| \sim \mathcal{O}(\xi)$

$$\frac{1}{r^{D-4+\bar{\eta}}} f_{dis}\left(\frac{r}{\xi}\right) \propto \int_1^{\infty} dL\, L^{1-D/2} e^{-\frac{L}{\xi^2} - \frac{r^2}{4L}}, \quad [58]$$

where $\bar{\eta} = 0$ in the mean-field approximation. Changing the integration variable from $L$ to $L' = L/r^2$ we have

$$\frac{1}{r^{D-4}} f_{dis}\left(\frac{r}{\xi}\right) \propto \frac{1}{r^{D-4}} \int_{1/r^2}^{\infty} dL'\, L'^{1-D/2} e^{-\frac{r^2}{\xi^2}L' - \frac{1}{4L'}}, \quad [59]$$

so that, since this scaling is valid for large $r$ we set the lower integration limit to 0 and we get

$$f_{dis}(x) \propto \int_0^{\infty} dL'\, L'^{1-D/2} e^{-x^2 L' - \frac{1}{4L'}} \propto x^{D/2-2} K_{D/2-2}(x), \quad [60]$$

where $K_\nu(x)$ is the modified Bessel function of the second kind. Note that as expected $f_{dis}(0)$ is finite due to the divergence of the Bessel function at small argument $K_\alpha(x) \propto x^{-\alpha}$. Similarly, to express the scaling function $f_{con}$, we perform the same computation using $P_L^{J\neq 0}(J)$ instead of $P_L^{J=0}$. In this case, the scaling function will have a second argument.

$$\frac{1}{r^{D-4+\bar{\eta}-\theta}} f_{con}\left(\frac{r}{\xi_{con}}, \frac{|J|}{t^{-\nu\theta}}\right) \propto \int_1^{\infty} dL\, L^{2-D/2} e^{-\frac{L}{\xi_{con}^2} - \frac{r^2}{4L}}, \quad [61]$$

where $\theta = 2$ in the mean-field approximation. Performing the same change of integration variable the result is

$$f_{con}(x,y) \propto \int_0^{\infty} dL'\, L'^{2-D/2} e^{-x^2 L' - \frac{1}{4L'}} \propto x^{D/2-3} K_{D/2-3}(x). \quad [62]$$

Note that at first order in $1/M$ $f_{con}(x,y)$ has no dependence on the second argument. As for $f_{dis}(0)$ we see that $f_{con}(0,y)$ is finite due to the small argument divergence of the Bessel function $K_\alpha(x) \propto x^{-\alpha}$.

$\mathcal{P}_2(x_1,x_2;\Delta E)$ can be expressed in terms of $P_2(x_1,x_2;u_1,u_2,J)$. According to the definitions given in the main text, the condition that the two spins are on the same cluster is that $R_{12} = R_{21} = 1$. This implies that $|u_1| < |J|$ and $|u_2| < |J|$. In this case, the local fields on the two spins are respectively $u_1 + \text{sign}(J)u_2$ and $u_2 + \text{sign}(J)u_1$ leading to $\Delta E = 2|u_1 + \text{sign}(J)u_2|$. This leads to the following exact expression:

$$\mathcal{P}_2(x_1,x_2;\Delta E) = \int_{-\infty}^{\infty} dJ \prod_{i=1,2} \int_{-|J|}^{|J|} du_i\, P_2(x_1,x_2;u_1,u_2,J)\, \delta\left(|u_1 + \text{sign}(J)\,u_2| - \frac{\Delta E}{2}\right). \quad [63]$$

Taking the Fourier transform we can plug the $1/M$ expression of $P_2(x_1, x_2; u_1, u_2, J)$ into the above equation:

$$FT[\mathcal{P}_2(x_1, x_2; \Delta E)] \propto \delta(k+k') \int_{-\infty}^{\infty} dJ \prod_{i=1,2} \int_{-|J|}^{|J|} du_i\, \widehat{P}_1(u_1)\widehat{P}_1(u_2) \frac{1}{(k^2+\tau+\rho|J|)^3} \delta\left(|u_1+\text{sign}(J)\,u_2| - \frac{\Delta E}{2}\right). \quad [64]$$

Note that we have selected only the contribution coming from $J \neq 0$ due to the constraints $|u_1| < |J|$ and $|u_2| < |J|$. In the large-distance limit $k \ll 1$ and in the critical region $\tau \ll 1$, the integral is dominated by $|J| \ll 1$, thus we will also have $|u_1| \ll 1$ and $|u_2| \ll 1$ because of $|u_1| < |J|$ and $|u_2| < |J|$. Therefore we can replace both $\widehat{P}_1(u_1)$ and $\widehat{P}_1(u_2)$ with $\widehat{P}_1(0)$. The integrals over $u_1$ and $u_2$ with the conditions $|u_1| < |J|$, $|u_2| < |J|$ and $|u_1 + u_2 \text{sign} J| = \Delta E$ leads to a factor $2|J| - \Delta E/2$ we than obtain the expression given in the main text:

$$FT[\mathcal{P}_2(x_1, x_2; \Delta E)] \propto \delta(k+k')\, \widehat{P}_1^2(0) \int_{\Delta E/4}^{\infty} dJ \frac{|J| - \Delta E/4}{(k^2+\tau+\rho|J|)^3} =$$

$$\delta(k+k')\, \widehat{P}_1^2(0) \int_0^{\infty} dJ \frac{J}{(k^2+\tau+\rho\Delta E/4+\rho J)^3} \propto \delta(k+k')\, \widehat{P}_1^2(0) \frac{1}{k^2+\tau+\rho\Delta E/4}. \quad [65]$$

We now want to generalize the above result to the case of a generic $f_{con}(x, y)$ in order to obtain the scaling function $f_2(\varrho, e)$ defined in the main text through:

$$\mathcal{P}_2(x_1, x_2; \Delta E) \equiv \frac{\widehat{P}_1^2(0)}{r^{D-4+\bar{\eta}+\theta}} f_2\left(\frac{r}{\xi}, \frac{\Delta E}{t^{\nu\theta}}\right). \quad [66]$$

We first observe that at leading order in $1/M$ we have $\bar{\eta} = 0$ and $\theta = 2$ and the scaling function $f_2(x, y)$ is

$$f_2(x, y) \propto \int_0^{\infty} dL \frac{e^{-Lx^2 - 1/(4L)}}{L^{D/2}} \propto x^{D/2-1} K_{D/2-1}(x). \quad [67]$$

The above result is the Fourier transform of Eq. (65). Note that, much as $f_{con}(x, y)$, $f_2(x, y)$ also does not depend on $y$ at this order. To obtain the general form of $f_2(x, y)$ it is convenient to switch from the scaling variables $r/\xi_{con}$ and $|J|/t^{\nu\theta}$ to the scaling variables $r/t^{-\nu}$ and $J/r^{-\theta}$. Indeed $r/\xi_{con}$ depends on $J$ through $\xi_{con} = t^{-\nu} f_\xi(|J|/t^{\nu\theta})$ and due to the integral over $J$ in expression (63) it is simpler to have a single scaling variable that is proportional to $J$. We thus introduce a new function $\tilde{f}_{con}\left(\frac{r}{t^{-\nu}}, \frac{|J|}{r^{-\theta}}\right)$ through

$$\tilde{f}_{con}\left(\frac{r}{t^{-\nu}}, \frac{|J|}{r^{-\theta}}\right) \equiv f_{con}\left(\frac{r}{t^{-\nu}} \bigg/ f_\xi\left(\frac{|J|}{r^{-\theta}}\frac{r^{-\theta}}{t^{\nu\theta}}\right), \frac{|J|}{r^{-\theta}}\frac{r^{-\theta}}{t^{\nu\theta}}\right). \quad [68]$$

Performing similar steps as those of Eq. (65) we obtain:

$$\mathcal{P}_2(x_1, x_2; \Delta E) \propto \frac{\widehat{P}_1^2(0)}{r^{D-4+\bar{\eta}+\theta}} \int_{\frac{\Delta E}{4r^{-\theta}}}^{\infty} d\widehat{J} \left(\widehat{J} - \frac{\Delta E}{4r^{-\theta}}\right) \tilde{f}_{con}\left(\frac{r}{t^{-\nu}}, \widehat{J}\right) \propto$$

$$\frac{\widehat{P}_1^2(0)}{r^{D-4+\bar{\eta}+\theta}} \int_0^{\infty} dy'\, y'\, \tilde{f}_{con}(r/t^{-\nu}, y' + \Delta E/(4r^{-\theta})) \equiv \frac{\widehat{P}_1^2(0)}{r^{D-4+\bar{\eta}+\theta}} \hat{f}_2(r/t^{-\nu}, \Delta E/r^{-\theta}) \quad [69]$$

leading to:

$$\hat{f}_2(x, y) \propto \int_0^{\infty} dy'\, y'\, \tilde{f}_{con}(x, y+y'). \quad [70]$$

From the scaling function $\hat{f}_2(r/t^{-\nu}, \Delta E/r^{-\theta})$ we can obtain the scaling function $f_2(r/\xi, \Delta E/t^{\nu\theta})$ by means of a change of scaling variables analogous to the one performed in Eq. (68).

The last scaling expression to be derived is the one in Eq. (12) of the main text

$$\sum_{x_2} \mathcal{P}_2(x_1, x_2, \Delta E) \propto \xi^{2-\eta} \tilde{f}_2\left(\frac{\Delta E}{t^{\nu\theta}}\right) \quad [71]$$

where $\eta = 0$ in the mean-field approximation. From Eq. (66) and $-\bar{\eta} + \eta = \theta - 2$ we obtain:

$$\tilde{f}_2(e) \propto \int_0^{\infty} \varrho^{1-\eta} f_2(\varrho, e)\, d\varrho. \quad [72]$$

**C. Observables on the Bethe Lattice.** At this point we can make use of the *Ansatz* distribution to compute the observables of interest on the Bethe lattice, that is the starting point of the $M$-layer expansion in power of $1/M$. In particular, we will denote with

$$\mathcal{C}_2(\mathcal{G}; L), \qquad \mathcal{D}_2(\mathcal{G}; L) \qquad \text{and} \qquad \mathcal{C}_3(\mathcal{G}; L). \quad [73]$$

respectively the two-point connected and disconnected functions and the three-point connected function on generic *diagrams* $\mathcal{G}$, as defined in the main text. On the Bethe lattice, neglecting long loops, two sites can only be connected by a unique sequence of adjacent edges, thus the two-point functions are computed on a line and the three-point function on the loop-less topology connecting three sites, respectively represented by $\mathcal{G}_1$ and $\mathcal{G}_3$ depicted in figures S1 and S2. The notation for $\mathcal{C}_2$, $\mathcal{D}_2$ and $\mathcal{C}_3$ will be particularly useful in the following where we compute corrections to the (mean-field) Bethe lattice results by means of the $M$-layer construction, considering the loop diagrams in figures S1 and S2.

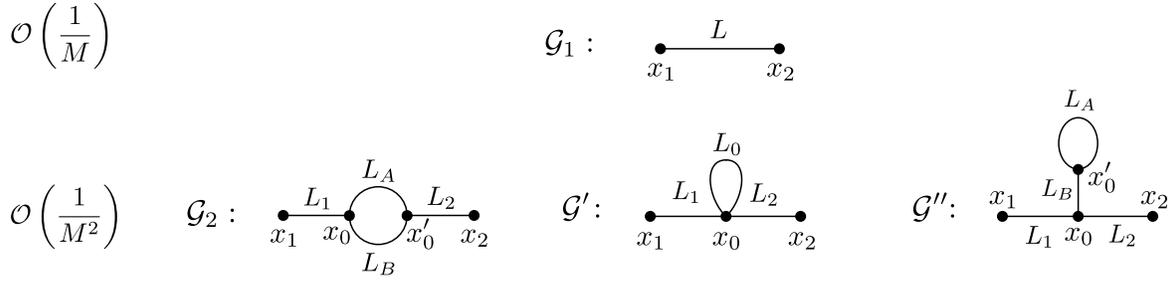

**Fig. S1.** Diagrams considered for the evaluation of two-point correlation functions (both connected and disconnected) up to one-loop order with the indicated explicit order in $1/M$ of their corresponding contribution.

### C.1. Two-point connected and disconnected functions.

In order to compute the two-point functions, $\mathcal{C}_2$ and $\mathcal{D}_2$, we recall their operational definitions, given in the main text. The disconnected function, on a line of length $L$, is simply defined to be the "non-trivial" $J=0$ part of the distribution $Q_L(u_1, u_2, J)$, setting the external effective fields to zero. On the other hand, the connected function can be obtained from the *Ansatz*, integrating with the conditions that the absolute values of the two effective fields at the extremities are lower than the absolute value of the effective coupling, together with the fact that the local field on the two sites is equal to the fixed value $\Delta E/2$, where we set $\Delta E = 0$, in order to take into account soft clusters, as described in the main text. Notice that the first two conditions make the two $J=0$ terms of the *Ansatz* irrelevant for the computation of the connected function since their contribution will be null. We will use the function $Q_L(h_1, h_2, J)$ defined in Eq. (50):

$$Q_L(h_1, h_2, J) = \delta(J)\,\widehat{P}_1(h_1)\widehat{P}_1(h_2) - P_L^{J=0}\,\delta(J)\,\widehat{P}_1(h_1)\widehat{P}_1(h_2) + P_L^{J\neq 0}(J)\,\widehat{P}_1(h_1)\widehat{P}_1(h_2),\qquad [74]$$

where we have introduced the following definitions:

$$P_L^{J=0} \equiv 2\,a\,L\,\lambda^L \qquad [75]$$

and

$$P_L^{J\neq 0}(J) \equiv a\,L^2\,\lambda^L\,\rho\,e^{-\rho|J|L}.\qquad [76]$$

Now, the disconnected function is the second term of Eq. (74) with $h_1 = 0 = h_2$

$$\mathcal{D}_2(\mathcal{G}_1; L) = -2a\,L\lambda^L\left(\widehat{P}_1(0)\right)^2.\qquad [77]$$

Conversely, to compute the connected function we focus on the last term, the only one that gives a non-zero contribution. Integrating over the three parameters $h_1$, $h_2$, and $J$ with the three conditions given by its definition we have the leading contribution

$$\mathcal{C}_2(\mathcal{G}_1; L) = \int dJ \int dh_1 \int dh_2 \prod_{i=1,2} \Theta(|J| - |h_i|)\,\delta(|u_1 + u_2\,\mathrm{sign}(J)|)\,P_L^{J\neq 0}(J)\,\widehat{P}_1(h_1)\widehat{P}_1(h_2) =$$

$$= \frac{4a}{\rho}\left(\widehat{P}_1(0)\right)^2 \lambda^L + \mathcal{O}\left(\frac{1}{L}\lambda^L\right),\qquad [78]$$

where, to compute the integrals we neglected $\mathcal{O}\left(\lambda^L/L\right)$ terms, that we cannot control with our approximate *Ansatz* distribution. All these computations are done neglecting higher order terms in the large-lengths limit, in the following we will not explicitly write all the corrections so that we have

$$\mathcal{C}_2(\mathcal{G}_1; L) = \frac{4a}{\rho}\left(\widehat{P}_1(0)\right)^2 \lambda^L.\qquad [79]$$

### C.2. Three-point connected function.

Here we compute the leading contribution to the three-point connected function, that comes from the diagram $\mathcal{G}_3$ in Fig. S2.

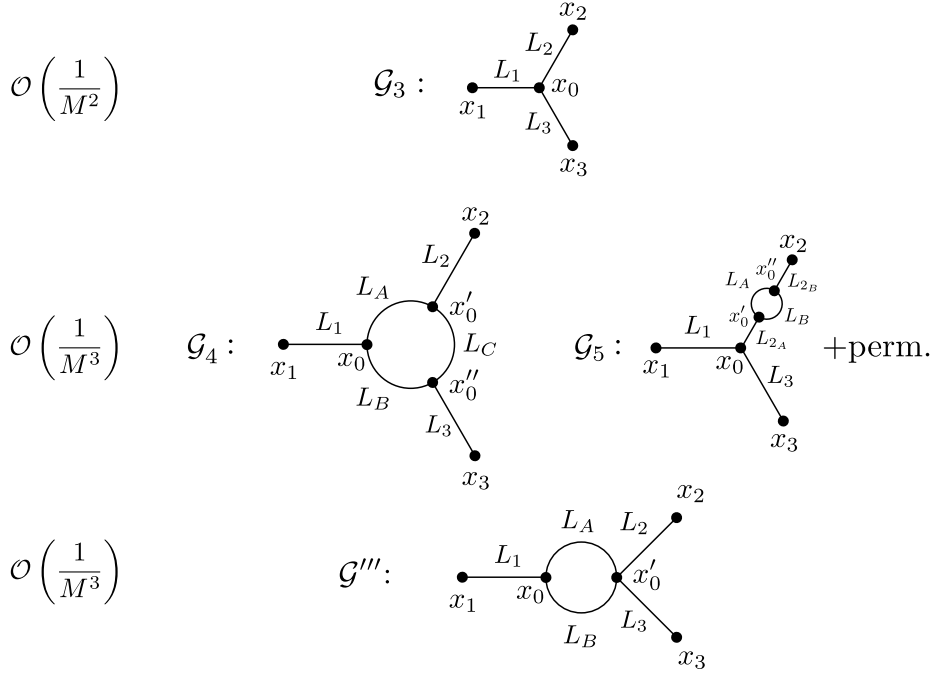

**Fig. S2.** Diagrams considered for the evaluation of the three-point connected correlation function up to one-loop order with the indicated explicit order in $1/M$ of their corresponding contribution.

In this case, 7 effective parameters can be identified, that constitute the following effective Hamiltonian
$$\mathcal{H}_{\mathcal{G}_3} = -h_1\sigma_1 - J_1\sigma_1\sigma_0 - h_2\sigma_2 - J_2\sigma_2\sigma_0 - h_3\sigma_3 - J_3\sigma_3\sigma_0 - h_0\sigma_0 \,. \qquad [80]$$

The observable we are interested in is the probability that $\sigma_1$, $\sigma_2$ and $\sigma_3$ belong to the same soft cluster. In other words, we want to compute the probability that the configuration of the three spins in the first excited state is flipped with respect to the one of the ground state and the energy difference is zero. This situation can be achieved only if also the central spin, $\sigma_0$ is flipped in the first excited state, with respect to its configuration in the ground state. The quantity to be averaged over the 7 parameters of the Hamiltonian of Eq. (80) is
$$\Theta(|J_1| - |h_1|)\Theta(|J_2| - |h_2|)\Theta(|J_3| - |h_{\to 0}|)\,\delta(h_{\to 3})\,, \qquad [81]$$
where
$$h_{\to 0} \equiv h_0 + \text{sign}(J_1)h_1 + \text{sign}(J_2)h_2 \qquad [82]$$
and
$$h_{\to 3} \equiv h_3 + \text{sign}(h_0 + \text{sign}(J_1)h_1 + \text{sign}(J_2)h_2)J_3 \,. \qquad [83]$$

In this case, the first two Heaviside step functions ensure that $\sigma_1$ and $\sigma_2$ are susceptible to a perturbation of $\sigma_0$, the third step function ensures that $\sigma_0$ is susceptible to a perturbation of $\sigma_3$ and the Dirac delta function ensures that the energy difference with respect to the ground state is zero. At this point, we need to evaluate the integral over the distributions of the triplets of the three lines composing the diagram. It can be simplified considering that the $J=0$ part of the distributions don't contribute to the integral, so that we can write, after the insertions of $2D-3$ cavity fields on $\sigma_0$ and $2D-1$ on $\sigma_1$, $\sigma_2$ and $\sigma_3$
$$\mathcal{C}_3(\mathcal{G}_3; L_1, L_2, L_3) = \int \Theta(|J_1| - |h_1|)\Theta(|J_2| - |h_2|)\Theta(|J_3| - |h_{\to 0}|)\,\delta(h_{\to 3})\,\widehat{P}_3(h_0) \prod_{i=1,2,3} P_{L_i}^{J\neq 0}(J_i)\,\widehat{P}_1(h_i)\,dJ_i\,dh_i\,, \qquad [84]$$

that is, neglecting $\mathcal{O}(\lambda^{3L}/L)$ terms (in the limit $L_1 = L_2 = L_3 \to L \gg 1$)
$$\mathcal{C}_3(\mathcal{G}_3; L_1, L_2, L_3) = \frac{64a^3}{\rho^3}\left(\widehat{P}_1(0)\right)^3 \widehat{P}_3(0)\lambda^{L_1+L_2+L_3}\,. \qquad [85]$$

Interestingly we notice that, apart from the factor $\widehat{P}_3(0)$ that comes from the internal vertex, the whole contribution can be factorized in the three contributions of the (external) lines:
$$\frac{4a}{\rho}\widehat{P}_1(0)\lambda^{L_i} \qquad \text{for } i = 1, 2, 3. \qquad [86]$$

This is a general feature of connected functions. To illustrate, consider $\mathcal{G}_3$ in Fig. S2. One of the external spins, say $\sigma_1$, belongs to the same soft cluster as the other external spins if and only if it is susceptible to the flip of the internal spin $\sigma_0$, *i.e.* $\sigma_1$ must flip from its ground state configuration along with $\sigma_0$. This occurs when the absolute value of the effective coupling between these two spins exceeds the local field acting on $\sigma_1$. Imposing this condition with the integration of the *Ansatz* for the corresponding external line of length $L_1$ yields the factor in Eq. (86).

In the following, we will denote with $\mathcal{G}^{amp}$ the *amputated* version of the generic diagram $\mathcal{G}$, obtained pruning all the external lines. In the case of a generic amputated diagram for the connected function, the contribution of the corresponding non-amputated diagram can be obtained by simply multiplying this factor for each external line, with the corresponding length. Notice that, given a generic diagram, when amputating an external line the degree, say $n$, of the vertex to which the leg is connected is decreased by 1. This should imply that the factor $\widehat{P}_n(0)$ of the non-amputated diagram becomes $\widehat{P}_{n-1}$ for the amputated one, since in the latter case $2D - (n-1)$ cavity fields,

$u \sim P_B(u)$, should be inserted in the vertex, while in the former case one of these fields is drawn from $g(u)$, coming from the external line. To simplify the computations of the following sections, we will instead consider the amputated diagrams as if one field is drawn from $g(u)$, coming from the (amputated) external leg. In this way the factor $\widehat{P}_n$ is present for the non-amputated as well as for the amputated version of the same diagram.

## 2. Application of the $M$-layer construction to the model

Here we explicitly apply the $M$-layer construction to the Edwards-Anderson spin glass model at zero temperature in an external field. In order to show how to get the expressions of $\chi_2$, $\chi_2^{dis}$ and $\chi_3$ given by Eqs. (28), (29) and (30) in the main text, in Sec. A we apply the generic rules of the $M$-layer construction, in Sec. B we complete the computation of the observables with loop diagrams and finally, in Sec. C, we arrive at the expressions of the susceptibilities on the $M$-layer lattice.

**A. Generalities of the construction.** Following the prescriptions of the $M$-layer construction as exposed in the section Material and Methods of the main text, and in ref. (29, 33, 34), we write here the expressions for the three observables defined in the main text, $\mathcal{P}_2(x_1, x_2, 0)$, $P^{dis}(x_1, x_2, 0, 0)$ and $\mathcal{P}_3(x_1, x_2, x_3, 0)$, taking into account the leading order and the one-loop corrections. In this SI, to avoid all the null inputs of the three observables we define the two-point connected and disconnected functions together with the three-point connected function respectively as

$$\mathcal{P}_2(x_1, x_2; 0) \equiv C_2(x_1, x_2), \qquad P^{dis}(x_1, x_2; 0, 0) \equiv D_2(x_1, x_2) \qquad \text{and} \qquad \mathcal{P}_3(x_1, x_2, x_3; 0) \equiv C_3(x_1, x_2, x_3), \qquad [87]$$

averaged over the disorder and the rewirings on the $M$-layer lattice. For each observable we will follow the prescribed steps: i) identify relevant diagrams; ii) compute $W$, $\mathcal{N}$ and $S$ factors; iii) sum the contributions and iv) compute the observable on the identified diagrams. The last step is the only model-dependent one and we leave it for the following Sec. B, while the other steps are analyzed here. We start with the connected and disconnected two-point functions, $C_2(x_1, x_2)$ and $D_2(x_1, x_2)$ together, because the contributing diagrams will be the same.

**Observables:** $C_2(x_1, x_2)$ **and** $D_2(x_1, x_2)$

❶ Identification of relevant diagrams

The simplest diagram connecting two points is the bare line, denoted as $\mathcal{G}_1$. When including the possibility of a loop, we consider the diagram composed of four lines and two degree-three vertices, where the two internal lines form a loop. We will refer to this diagram as $\mathcal{G}_2$ (see Fig. S1).

In principle, two additional diagrams could contribute to two-point observables: the *tadpole-type* diagrams $\mathcal{G}'$ and $\mathcal{G}''$, depicted in Fig. S1. When computing observables on given diagrams, Sec. B, we prove that these contributions are negligible in the large-length limit considered in this work. Specifically, we show why the contributions to both the two-point disconnected and connected functions vanish (up to negligible corrections) when computed on the diagrams $\mathcal{G}'$ and $\mathcal{G}''$. The argument leads to a more general conclusion: the insertion of a tadpole-type diagram on any generic line is negligible for the computations presented here. This observation allows us to disregard such diagrams even when they are attached to a generic $k$-point function, and in particular, for the three-point function, as we will illustrate below.

In the following, we focus exclusively on $\mathcal{G}_1$ and $\mathcal{G}_2$.

❷ $W$, $\mathcal{N}$ and $S$ factors

Given a generic diagram $\mathcal{G}$ with lengths $\vec{L}$ and external vertices at $x_1, ..., x_q$, to compute $\mathcal{N}(\mathcal{G}; \vec{L}; x_1, ..., x_q)$, for each internal vertex we have to ensure that the directions of the entering lines are different. This requires considering the number of NBP between two points with assigned incoming and outcoming directions. However, for large $L$, the result becomes independent of the direction and it is then given by $\mathcal{N}_L(x, y)/(2D)^2$. Thus, to compute $\mathcal{N}(\mathcal{G}; \vec{L}; x_1, ..., x_q)$ we have to multiply for each line by a factor $\mathcal{N}_L(x, y)/(2D)^2$. Then, to take into account the sum over allowed directions at each vertex, we additionally multiply by a factor $2D$ for each external vertex and a factor $\frac{(2D)!}{(2D-3)!}$ for each internal cubic vertex. The factors $W$, $\mathcal{N}$ and $S$ result to be the following for the two-point diagrams:

Diagram $\mathcal{G}_1$

- $W(\mathcal{G}_1) = \frac{1}{M}$;
- $\mathcal{N}(\mathcal{G}_1; L; x_1, x_2) = \mathcal{N}_L(x_1, x_2)$;
- $S(\mathcal{G}_1) = 1$.

Diagram $\mathcal{G}_2$

- $W(\mathcal{G}_2) = \frac{1}{M^2}$
- $\mathcal{N}(\mathcal{G}_2; \vec{L}; x_1, x_2) = (2D)^2 \left( \frac{(2D)!}{(2D-3)!} \right)^2 \sum_{x_0, x'_0} \frac{\mathcal{N}_{L_1}(x_1, x_0)}{(2D)^2} \frac{\mathcal{N}_{L_A}(x_0, x'_0)}{(2D)^2} \frac{\mathcal{N}_{L_B}(x_0, x'_0)}{(2D)^2} \frac{\mathcal{N}_{L_2}(x'_0, x_2)}{(2D)^2}$;
- $S(\mathcal{G}_2) = \frac{1}{2}$.

❸ Sum of the contributions

The perturbative expression of the two-point connected function on the $M$-layer lattice, averaged over the rewirings, $C_2(x_1, x_2)$, is the sum over the two relevant diagrams, up to one-loop contributions:

$$C_2(x_1, x_2) = \frac{1}{M} \sum_L \mathcal{N}_L(x_1, x_2) \mathcal{C}_2^{lc}(\mathcal{G}_1; L) + \frac{1}{2M^2} \sum_{\vec{L}} \mathcal{N}(\mathcal{G}_2; \vec{L}; x_1, x_2) \mathcal{C}_2^{lc}(\mathcal{G}_2; \{\vec{L}\}) + \mathcal{O}\left(\frac{1}{M^3}\right), \qquad [88]$$

where $\vec{L} = (L_1, L_A, L_B, L_2)$. Analogously we can write the same expansion for the disconnected function

$$D_2(x_1, x_2) = \frac{1}{M} \sum_L \mathcal{N}_L(x_1, x_2) \mathcal{D}_2^{lc}(\mathcal{G}_1; L) + \frac{1}{2M^2} \sum_{\vec{L}} \mathcal{N}(\mathcal{G}_2; \vec{L}; x_1, x_2) \mathcal{D}_2^{lc}(\mathcal{G}_2; \{\vec{L}\}) + \mathcal{O}\left(\frac{1}{M^3}\right), \qquad [89]$$

where again we remark that the only difference is the observable computed on a given diagram.

Notice also that we generalized the notation for the observables on the Bethe lattice, including the superscript "lc"

$$\mathcal{C}_2^{lc}(\mathcal{G}; \{\vec{L}\}) \qquad \text{and} \qquad \mathcal{D}_2^{lc}(\mathcal{G}; \{\vec{L}\}), \qquad [90]$$

which stands for "line-connected", a definition that allows to correctly isolate the loop contributions, as prescribed by the $M$-layer construction. As explained in the Material and Methods section of the main text, loop corrections to observables include the contribution coming from the corresponding *loop-less* sub-diagram. To avoid counting twice these contributions, the $M$-layer construction requires the computation of "line-connected" observables. The general definition can be found in ref. (29), while for the diagrams considered in this paper it is sufficient to state that to compute a generic *line-connected* observable on a diagram $\mathcal{G}$, $O^{lc}(\mathcal{G})$, one has to compute the observable on $\mathcal{G}$ and then subtract all the contributions from $O$ computed on diagrams where a line composing the loop (if any) is removed. It follows that a generic observable computed on a loop-less diagram (e.g. $\mathcal{G}_1$) is also *line-connected*: $O(\mathcal{G}_1) = O^{lc}(\mathcal{G}_1)$. The same notation is used in the following for the three-point function: $\mathcal{C}_3^{lc}(\mathcal{G}; \{\vec{L}\})$.

**Observable:** $C_3(x_1, x_2, x_3)$

❶ Identification of relevant diagrams

The simplest diagram connecting three points is the bare degree-three vertex, which we denote as $\mathcal{G}_3$. To account for the possibility of a loop, we consider a diagram composed of six lines and three degree-three vertices, where the three internal lines form a loop. This diagram is referred to as $\mathcal{G}_4$. At the one-loop level, three additional diagrams connect three points with a single loop. These are similar to $\mathcal{G}_3$, but with one of the external legs dressed by $\mathcal{G}_2$. We refer to such a diagram as $\mathcal{G}_5$. All these diagrams are shown in Fig. S2.

Other possible diagrams are obtained by setting the length of one line to zero (e.g., the diagram $\mathcal{G}'''$ in Fig. S2). These diagrams would, in principle, contribute as additional diagrams under the prescriptions of the $M$-layer construction. This is because setting one length to zero increases the number of lines incident to a vertex, thereby altering the combinatorial factors associated with the number of NBP. However, in Sec. B, we provide an argument explaining why such *sub-diagrams* can be neglected in our computations.

❷ $W$, $\mathcal{N}$ and $S$ factors

Diagram $\mathcal{G}_3$

- $W(\mathcal{G}_3) = \frac{1}{M^2}$;
- $\mathcal{N}(\mathcal{G}_3; L_1, L_2, L_3; x_1, x_2, x_3) = (2D)^3 \frac{(2D)!}{(2D-3)!} \sum_{x_0} \prod_{i=i}^3 \frac{\mathcal{N}_{L_i}(x_i, x_0)}{(2D)^2}$;
- $S(\mathcal{G}_3) = 1$.

Diagram $\mathcal{G}_4$

- $W(\mathcal{G}_4) = \frac{1}{M^3}$;
- $\mathcal{N}(\mathcal{G}_4; \vec{L}'; x_1, x_2, x_3) = (2D)^3 \left(\frac{(2D)!}{(2D-3)!}\right)^3 \sum_{x_0, x_0', x_0''} \frac{\mathcal{N}_{L_1}(x_1, x_0)}{(2D)^2} \frac{\mathcal{N}_{L_2}(x_2, x_0')}{(2D)^2} \frac{\mathcal{N}_{L_3}(x_3, x_0'')}{(2D)^2} \frac{\mathcal{N}_{L_A}(x_0, x_0')}{(2D)^2} \frac{\mathcal{N}_{L_B}(x_0, x_0'')}{(2D)^2} \frac{\mathcal{N}_{L_C}(x_0', x_0'')}{(2D)^2}$;
- $S(\mathcal{G}_4) = 1$.

Diagram $\mathcal{G}_5$

- $W(\mathcal{G}_5) = \frac{1}{M^3}$
- $\mathcal{N}(\mathcal{G}_5; \vec{L}''; x_1, x_2, x_3) = (2D)^3 \left(\frac{(2D)!}{(2D-3)!}\right)^3 \sum_{x_0, x_0', x_0''} \frac{\mathcal{N}_{L_1}(x_1, x_0)}{(2D)^2} \frac{\mathcal{N}_{L_{2_A}}(x_0, x_0')}{(2D)^2} \frac{\mathcal{N}_{L_{2_B}}(x_2, x_0'')}{(2D)^2} \frac{\mathcal{N}_{L_3}(x_3, x_0)}{(2D)^2} \prod_{l = L_A, L_B} \frac{\mathcal{N}_l(x_0', x_0'')}{(2D)^2}$;
- $S(\mathcal{G}_5) = \frac{1}{2}$.

❸ Sum of the contributions

As for the two-point function, we can write the perturbative expression of the three-point connected function on the $M$-layer lattice, averaged over the rewirings:

$$C_3(x_1, x_2, x_3) = \frac{1}{M^2} \sum_{L_1, L_2, L_3} \mathcal{N}(\mathcal{G}_3; L_1, L_2, L_3; x_1, x_2, x_3) \mathcal{C}_3^{lc}(\mathcal{G}_3; \{L_1, L_2, L_3\}) +$$

$$+ \frac{1}{M^3} \sum_{\vec{L}'} \mathcal{N}(\mathcal{G}_4; \vec{L}'; x_1, x_2, x_3) \mathcal{C}_3^{lc}(\mathcal{G}_4; \{\vec{L}'\}) +$$

$$+ \frac{1}{2M^3} \sum_{\vec{L}''} \mathcal{N}(\mathcal{G}_5; \vec{L}''; x_1, x_2, x_3) \mathcal{C}_3^{lc}(\mathcal{G}_5; \{\vec{L}''\}) + \mathcal{O}\left(\frac{1}{M^4}\right), \qquad [91]$$

where $\vec{L}' = (L_1, L_2, L_3, L_A, L_B, L_C)$ and $\vec{L}'' = (L_1, L_{2_A}, L_A, L_B, L_{2_B}, L_3)$.

The last step of the $M$-layer construction is the actual computation of the observables on the given diagrams, to be done in the next section.

**B. One-loop corrections of the observables on given diagrams.** We already computed the contributions of the observables on the Bethe lattice in Sec. 1, which corresponds to the leading order in the $M$-layer framework, in this section we will perform the computation on the remaining topologies: $\mathcal{G}_2$ for the two-point functions, $\mathcal{G}_4$ and $\mathcal{G}_5$ for the three-point function. Notice that these "diagrams" represent a Bethe lattice in which, eventually, loops are properly added in order to consider more complicated paths connecting the spins, resulting from the random rewiring. In this framework we can use Bethe lattice techniques, explained in Sec. 1, for the spin glass model in a field at zero temperature. We first compute the observables and then we explain why other possible diagrams are not relevant for our computations.

**Two-point functions** Now we want to compute the first corrections to $\mathcal{C}_2^{lc}$ and $\mathcal{D}_2^{lc}$. The corresponding diagram is $\mathcal{G}_2$, depicted in Fig. S1. The idea is to repeat the same steps done for $\mathcal{G}_1$, to do so we need to compute the effective triplet on the new topology given by the loop diagram. The first step is to add together two lines to compose the amputated loop, $\mathcal{G}_2^{amp}$. In this case we obtain a simple distribution from the two lines of lengths $L_A$ and $L_B$

$$P_{L_A, L_B}^{amp}(v_1, v_2, J) = \int dv_1^A \, dv_2^A \, dv_1^B \, dv_2^B \, dJ_A \, dJ_B \, \delta(v_1^A + v_1^B - v_1) \, \delta(v_2^A + v_2^B - v_2) \, \delta(J_A + J_B - J) \times$$
$$\times P_{L_A}(v_1^A, v_2^A, J_A) \, P_{L_B}(v_1^B, v_2^B, J_B). \quad [92]$$

The result is the distribution of the effective triplet of the amputated $\mathcal{G}_2$ diagram, in the large-lengths limit. At this point we should join the two external legs, using the rules of Tab. S1. The resulting distribution does not take into account the line-connected definition, which, in this case, requires subtracting the two loop-less "sub-diagram" given by $\mathcal{G}_2$ removing the upper or lower line composing the loop. To automatically subtract these two contributions one can simply remove the first term from the complete *Ansatz* of the two lines of the loop

$$P_{L_i}(v_1^i, v_2^i, J_i) \quad \rightarrow \quad P_{L_i}(v_1^i, v_2^i, J_i) - \delta(J) \, g(v_1^i) \, g(v_2^i) \qquad i = A, B, \quad [93]$$

while keeping the complete expression for the external lines: $P_{L_1}(u_1, v_1^1, J_1)$ and $P_{L_2}(v_2^2, u_2, J_2)$. In this way it is possible to figure out that the subtractions due to the line-connected definition are done, the only difference would be the subtraction of a contribution of a diagram in which the two lines of the loop are both removed, but for the observables we compute in this paper this contribution will be zero. The expression of the distribution of the effective triplet, once the external legs are added, is

$$P_{\{\vec{L}\}}^{lc}(u_1, u_2, J) = g(u_1)g(u_2) \left[ 64 \frac{a^4 \left(\widehat{P}_3(0)\right)^2}{\rho^2} \frac{\lambda^{L_1+L_2+L_A+L_B}}{L_A + L_B} L_A L_B (L_1 + L_2 + L_A + L_B) \delta(J) \right.$$
$$\left. - 32 \frac{a^4 \left(\widehat{P}_3(0)\right)^2}{\rho} \frac{\lambda^{L_1+L_2+L_A+L_B}}{L_A^2 - L_B^2} L_A L_B \, e^{-|J|(L_1+L_2)\rho} \left( e^{-|J|L_A\rho} L_A (L_1 + L_2 + L_A)^2 - e^{-|J|L_B\rho} L_B (L_1 + L_2 + L_B)^2 \right) \right], \quad [94]$$

where again $\vec{L} = (L_1, L_A, L_B, L_2)$. Before calculating the observables let us notice that the integral over the effective coupling again gives zero, as for the *Ansatz* distribution on $\mathcal{G}_1$. We expect this property to be general to every loop order for two-point diagrams.

Now we should perform the same steps done for $\mathcal{G}_1$ in order to compute the loop contributions for the two-point connected and disconnected functions. Once $2D - 1$ cavity fields are added to each end of the diagram, the disconnected function is simply the $J = 0$ term:

$$\mathcal{D}_2^{lc}(\mathcal{G}_2; \{\vec{L}\}) = 64 \frac{a^4 \left(\widehat{P}_3(0)\right)^2 \left(\widehat{P}_1(0)\right)^2}{\rho^2} \frac{\lambda^{L_1+L_2+L_A+L_B}}{L_A + L_B} L_A L_B (L_1 + L_2 + L_A + L_B). \quad [95]$$

To obtain the result for the connected function we also have to integrate over the effective fields and coupling with the conditions that the absolute value of the coupling is greater than the ones of the fields and the local field is zero. Finally we have

$$\mathcal{C}_2^{lc}(\mathcal{G}_2; \{\vec{L}\}) = -\frac{128 a^4}{\rho^3} \left(\widehat{P}_1(0)\right)^2 \left(\widehat{P}_3(0)\right)^2 \frac{L_A L_B}{L_A + L_B} \lambda^{L_1+L_A+L_B+L_2}. \quad [96]$$

Notice that the dependence on the lengths in this loop correction of the connected function is a non-trivial function of the internal line lengths $L_A$ and $L_B$, whereas the dependence on the external lines $L_1$ and $L_2$ factorizes in two exponentials $\lambda^{L_1}$ and $\lambda^{L_2}$, as observed in the comment after Eq. (86).

As a last remark we note that a generalization of the computation of $\mathcal{C}_2^{lc}(\mathcal{G}_2; \{\vec{L}\})$ is necessary to compute the three-point connected function on the loop diagram $\mathcal{G}_4$, this is what is done in the next Section.

**B.1. Three-point function.** Here we show how to compute the two one-loop contributions to the three-point connected function. We start from $\mathcal{G}_5$ which is easily done repeating the argument given above for the external lines: the contribution of the "dressed" external leg, that is $\mathcal{C}_2^{lc}(\mathcal{G}_2; \{\vec{L}\})$, can be multiplied to the contributions of the remaining two external lines, resulting in

$$\mathcal{C}_3^{lc}(\mathcal{G}_5; \vec{L}') = -\frac{2048 a^6}{\rho^5} \left(\widehat{P}_1(0)\right)^3 \left(\widehat{P}_3(0)\right)^3 \frac{L_A L_B}{L_A + L_B} \lambda^{L_1 + L_{2_A} + L_{2_B} + L_A + L_B + L_3}. \quad [97]$$

The last diagram to be computed is $\mathcal{G}_4$, to do so we generalize the procedure applied to the connected function on the two-point diagram $\mathcal{G}_2$. We will perform the computation in such a way that the contribution of a generic $n$-point connected function on an amputated loop diagram (*i.e.* a ring with $n$ vertices and $n$ lines) can be computed, then, as we argued above, the contributions of the external lines can be simply multiplied.

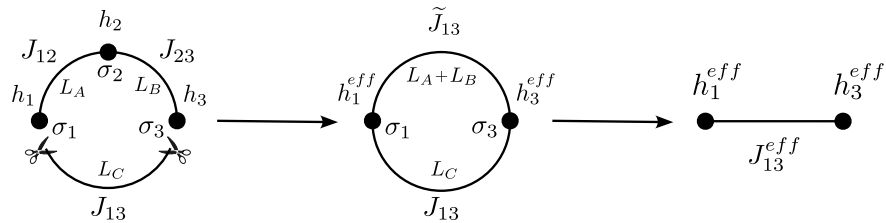

**Fig. S3.** Graphic representation of the procedure to compute $\mathcal{C}_3^{lc}(\mathcal{G}_4; \vec{L})$. Once the probability density that $\sigma_2$ is in the same soft cluster as $\sigma_1$ and $\sigma_2$ the loop systems can be treated as an effective two-spins system, from which it is simple to compute the connected correlation.

The goal is to compute the three-point connected function of an effective system given by the following Hamiltonian

$$\mathcal{H}_{\mathcal{G}_4^{amp}} = -h_1 \sigma_1 - J_{12} \sigma_1 \sigma_2 - h_2 \sigma_2 - J_{23} \sigma_2 \sigma_3 - h_3 \sigma_3 - J_{13} \sigma_1 \sigma_3, \quad [98]$$

with three effective fields on the three spins $\sigma_1$, $\sigma_2$ and $\sigma_3$ and three effective couplings. Notice that we denoted the amputated three-point loop diagram with $\mathcal{G}_4^{amp}$. The amputated diagram is obtained by removing as usual the external legs and we require that the three spins belong to the same soft cluster. As already explained, we define the amputated diagram assuming that the distribution of the fields coming from the external legs is given by $g(u)$ instead of $P_B(u)$.

In practice for each line of length $L$ we have a factor $g(u)g(u')(-\delta(J)P_L^{J=0} + P_L^{J\neq 0}(J))$ and for each internal vertex we have $2D-3$ additional incoming fields distributed as $P_B(u)$ and one incoming field distributed as $g(u)$ (according to our definition of amputated diagrams). Then we have to integrate this pseudo-measure over $h_1, h_2, h_3$ and $J_{12}, J_{23}, J_{13}$ with the condition that the three spins are on a soft cluster. Note that we use the expression pseudo-measure because the resulting product is not positive definite. Due to the factorized nature of the *Ansatz* the pseudo-measure is the sum of terms that are factorized over $h_1, h_2, h_3$ and $J_{12}, J_{23}, J_{13}$. In particular there is a factor $\widehat{P}_3(h_1)\widehat{P}_3(h_2)\widehat{P}_3(h_3)$ associated to $h_1, h_2, h_3$.

We will first consider the case in which all three couplings are different from zero and consider at the end the case in which one of them vanishes (if two or more couplings are zero the three spins cannot be in the same soft cluster). In this case, the factor is:

$$\widehat{P}_3(h_1)\widehat{P}_3(h_2)\widehat{P}_3(h_3) a^3 \rho^3 e^{-|J_{12}|L_A+|J_{23}|L_B+|J_{13}|L_C} L_A^2 L_B^2 L_C^2 \lambda^{L_A+L_B+L_C} . \quad [99]$$

To study the problem it is convenient to remove one line, say $L_C$, and then consider the effect of putting it back again in the system as depicted in Fig. S3, this amounts to assume that $J_{13}=0$. In this way, we reduce the three-spins loop to a simple line for which we know how to compute the correlation function. We can then compute the effective coupling between $\sigma_1$ and $\sigma_3$, let us call it $\widetilde{J}_{13}$ and the two effective fields $h_1^{eff}$ and $h_3^{eff}$. However, since $J_{12}$ and $J_{23}$ are small, the effect of integrating out $\sigma_2$ produces a negligible shift on the fields, $h_1^{eff} \approx h_1$ and $h_3^{eff} \approx h_3$. Therefore it is correct to consider that integrating out $h_2$, $J_{12}$ and $J_{23}$ only leads to an effective coupling $\widetilde{J}_{13}$ while the fields distributions remain $\widehat{P}_3(h_1)\widehat{P}_3(h_3)$. Without $L_C$, the condition for the three spins to be in the same cluster is equivalent to the condition that spins $\sigma_1$ and $\sigma_3$ are in a soft cluster. Indeed by flipping $\sigma_3$, $\sigma_1$ flips iff $\sigma_2$ flips. As we have seen repeatedly, the condition that spins $\sigma_1$ and $\sigma_3$ are in a soft cluster is simply that the absolute value of $\widetilde{J}_{13}$ is larger than the absolute values of the effective fields $h_1$ and $h_3$ and that $h_1 = -h_3$. According to Tab. S1 the case $\widetilde{J}_{13}=0$ can only occur if spin $\sigma_2$ is blocked or if either $J_{12}$ or $J_{23}$ vanish. Since we are considering the case in which the three couplings $J_{12}, J_{23}, J_{13}$ are different from zero and $\sigma_2$ is in the soft cluster it follows that $\widetilde{J}_{13} \neq 0$. To re-introduce $L_C$ now we can follow the argument used for the loop of diagram $\mathcal{G}_2$. In particular the effective coupling is $J_{13}^{eff} = \widetilde{J}_{13} + J_{13}$. We will now consider two different possibilities: i) in the first case the total effective coupling, $J_{13}^{eff} = \widetilde{J}_{13} + J_{13}$, between $\sigma_1$ and $\sigma_3$ has the same sign of $\widetilde{J}_{13}$, ii) in the second case instead it has the opposite sign.

*First case* - In the first case, adding $L_C$ does not change the behavior of the three spins: if the fields on $\sigma_1$ and $\sigma_3$, respectively $h_1$ and $h_3$, are smaller (in absolute value) than $J_{13}^{eff}$ and $h_1 + \text{sign}(J_{13}^{eff})h_3 = 0$ all three spins belong to the same cluster. From now on we consider the case $\widetilde{J}_{13} > 0$, the case $\widetilde{J}_{13} < 0$ can be accounted for by multiplying the final result by 2, for symmetry reasons. We can already compute the contribution to the three-point connected function of the first case: we have to join $L_A$ and $L_B$ as described above for the self-consistency of the *Ansatz* and then add the contribution of $L_C$ with the constraint $J_{13} > -\widetilde{J}_{13}$ that ensures $J_{13}^{eff} > 0$. Note however that, according to our definition of amputated diagrams, when we join $L_A$ and $L_B$ the field on $\sigma_2$ coming from the external line is not distributed as $P_B(u)$ but as $g(u)$, therefore, the distribution of $\widetilde{J}_{13}$ is *not* given by $(L_A+L_B)^2 a\lambda^{L_A+L_B}\rho \exp(-\rho(L_A+L_B)\widetilde{J}_{13})$ but there is an additional factor $4a\widehat{P}_3(0)/\rho$. Therefore the effective coupling is distributed as $\widetilde{P}_{L_A+L_B}^{J\neq 0}(\widetilde{J}_{13}) \equiv P_{L_A+L_B}^{J\neq 0}(\widetilde{J}_{13})4a\widehat{P}_3(0)/\rho$. The contribution of the first case is then:

$$2\int_0^\infty dJ_{13}^{eff} \int_0^\infty d\widetilde{J}_{13} \int_{-\widetilde{J}_{13}}^\infty dJ_{13} \int_{-\infty}^\infty dh_1 \int_{-\infty}^\infty dh_3 \, \widetilde{P}_{L_A+L_B}^{J\neq 0}(\widetilde{J}_{13}) P_{L_C}^{J\neq 0}(J_{13}) \widehat{P}_3(h_1) \widehat{P}_3(h_3) \times$$

$$\delta(J_{13}^{eff} - (\widetilde{J}_{13} + J_{13})) \prod_{i=1,3} \Theta(|J_{13}^{eff}| - |h_i|) \delta(h_1 + h_3 \text{sign}(J_{13}^{eff})) =$$

$$= \frac{16 a^3 \left(\widehat{P}_3(0)\right)^3}{\rho^2} \frac{L_A^2 + L_B^2 + 2L_C^2 + 2L_A L_B + 2L_B L_C + 2L_A L_C}{L_A + L_B + L_C} \lambda^{L_A+L_B+L_C} , \quad [100]$$

where the factor 2 is included to take into account the case $\widetilde{J}_{13} < 0$.

*Second case* - In the second case, after the addition of $J_{13}$ spin $\sigma_1$ and $\sigma_3$ are oriented differently in the soft cluster and *it may now happen that the spin $\sigma_2$ is no longer in the soft cluster*. Indeed spin $\sigma_2$ flips always when $\sigma_1$ and $\sigma_3$ switch between the two configurations satisfying $\sigma_1\sigma_3 = \widetilde{J}_{13}$ but not necessarily when $\sigma_1$ and $\sigma_3$ switch between the two configuration satisfying $\sigma_1\sigma_3 = -\widetilde{J}_{13}$. We define $P_{L_A,L_B}^{NB}(\widetilde{J}_{13})$ the probability density of having $\widetilde{J}_{13}$ constrained to this additional condition. The suffix $(NB)$ stands for *non-blocking* meaning that the central spin $\sigma_2$ is not blocked if $(\sigma_1,\sigma_3) \to (-\sigma_1,-\sigma_3)$ independently of $\widetilde{J}_{13}$. Note again that we are safely assuming that since the couplings are small the fields $h_1$ and $h_3$ are not changed. From now on we consider the case $\widetilde{J}_{13} > 0$, and the case $\widetilde{J}_{13} < 0$ can be accounted multiplying the result by 2 for symmetry reasons. In this case we have to select those instances of the disorder such that *without* the line $L_C$, spin $\sigma_2$ flips not only when $\sigma_1$ and $\sigma_3$ switch between $(1,1)$ and $(-1,-1)$ but *also* when they switch between $(1,-1)$ and $(-1,1)$. Let us again consider only positive couplings $J_{12}>0, J_{23}>0$, the negative case can be argued to be the same. Let us also consider the case $J_{12} > J_{23}$, once more the other case will give a similar result. In this case, the condition for $\sigma_2$ to be susceptible to the flip of $\sigma_1$ or $\sigma_3$ is $(J_{12} - J_{23}) > |h_2|$, because in this second case the line $L_C$ will change the sign of the coupling between $\sigma_1$ and $\sigma_3$ and they will be antiparallel in the ground state. One can see that, when this condition is met, the effective coupling $\widetilde{J}_{13}$ is the *minimum* between $J_{12}$ and $J_{23}$, as can be read from Tab. S1. The *Ansatz* for the line, of length $L_A + L_B$, joining the three spins is

$$\widehat{P}_3(h_2) \rho^2 a^2 L_A^2 L_B^2 \lambda^{L_A+L_B} e^{-\rho L_A J_{12}} e^{-\rho L_B J_{23}} . \quad [101]$$

Now, in order to obtain $P_{L_A,L_B}^{NB}(\widetilde{J}_{13} > 0)$, we have to integrate the previous expression with the condition $|h_2| < (J_{12} - J_{23})$, which also implies that the effective coupling is equal to the minimum coupling $\widetilde{J}_{13} = J_{23}$ (remember that we are considering $J_{12} > J_{23}$). We recall that we are neglecting the effect of the whole procedure on the distribution of $h_1$ and $h_3$, indeed according to Tab. S1 the field on $\sigma_1$ will not be changed after the integration over $h_2$, $J_{12}$ and $J_{13}$ while the field on $\sigma_3$ will get a contribution proportional to $J_{23}$ that can be

neglected with respect to $h_3$, since $L_B \gg 1$ implies $J_{23} \ll 1$. The resulting contribution for $J_{12} > 0$ and $J_{13} > 0$ and $J_{12} > J_{23}$ is then :

$$\rho^2 a^2 L_A^2 L_B^2 \lambda^{L_A+L_B} \widehat{P}_3(0) e^{-\rho L_B \widetilde{J}_{13}} \int_{\widetilde{J}_{13}}^{\infty} 2(J_{12} - \widetilde{J}_{13}) e^{-\rho L_A J_{12}} dJ_{12} = 2 a^2 L_B^2 \lambda^{L_A+L_B} \widehat{P}_3(0) e^{-\rho(L_B+L_A)\widetilde{J}_{13}} , \qquad [102]$$

where the integral over $h_2$ contributes a factor $2(J_{12} - \widetilde{J}_{13})\widehat{P}_3(0)$ because $|h_2| < (J_{12} - \widetilde{J}_{13}) \ll 1$. The case $J_{12} < J_{23}$ leads naturally to the same expression with $L_A \leftrightarrow L_B$, while the case $J_{12} < 0$ and $J_{23} < 0$ can be reduced to $J_{12} > 0$ and $J_{23} > 0$ by changing $\sigma_2 \to -\sigma_2$ and leads to an additional factor two. The final expression for $P_{L_A,L_B}^{NB}(\widetilde{J}_{13})$ is then:

$$P_{L_A,L_B}^{NB}(\widetilde{J}_{13}) = 4 a^2 \widehat{P}_3(0) (L_A^2 + L_B^2) \lambda^{L_A+L_B} e^{-\rho |\widetilde{J}_{13}|(L_A+L_B)} . \qquad [103]$$

We wrote the absolute value $|\widetilde{J}_{13}|$ as the above expression is valid also for $\widetilde{J}_{13} < 0$ for symmetry reasons. Now it's possible to re-introduce the line with coupling $J_{13}$, contributing the factor

$$P_{L_C}^{J\neq 0}(J_{13}) = a \rho \lambda^{L_C} L_C^2 e^{-\rho |J_{13}|L_C} , \qquad [104]$$

with the constraint that $J_{13} < -\widetilde{J}_{13}$. Adding a factor 2 to include the case $\widetilde{J}_{13} < 0$ we have the following contribution:

$$2 \int_{-\infty}^{0} dJ_{13}^{eff} \int_{0}^{\infty} d\widetilde{J}_{13} \int_{-\infty}^{-\widetilde{J}_{13}} dJ_{13} \int_{-\infty}^{\infty} dh_1 \int_{-\infty}^{\infty} dh_3 \, 4 a^2 \widehat{P}_3(0) (L_A^2 + L_B^2) \lambda^{L_A+L_B} e^{-\rho \widetilde{J}_{13}(L_A+L_B)} P_{L_c}^{J\neq 0}(J_{13}) \widehat{P}_3(h_1) \widehat{P}_3(h_3) \times$$

$$\delta(J_{13}^{eff} - (\widetilde{J}_{13} + J_{13})) \prod_{i=1,3} \Theta(|J_{13}^{eff}| - |h_i|) \delta(h_1 + h_3 \,\text{sign}(J_{13}^{eff})) = \frac{16 a^3 \left(\widehat{P}_3(0)\right)^3}{\rho^2} \frac{L_A^2 + L_B^2}{L_A + L_B + L_C} \lambda^{L_A+L_B+L_C} , \qquad [105]$$

and adding it to the first contribution, Eq. (100), we get:

$$\frac{32 a^3 \left(\widehat{P}_3(0)\right)^2}{\rho^2} \frac{L_A^2 + L_B^2 + L_C^2 + L_A L_B + L_A L_C + L_B L_C}{L_A + L_B + L_C} \lambda^{L_A+L_B+L_C} . \qquad [106]$$

It remains to be considered the case in which one of the lines has zero coupling. Let us consider the case in which $J_{13} = 0$, then the other cases are the same, up to a re-labeling of the lines. The pseudo-measure contribution is

$$-2 \widetilde{P}_3(h_1) \widetilde{P}_3(h_2) \widetilde{P}_3(h_3) a^3 \rho^2 \delta(J_{13}) e^{-|J_{12}|L_A + |J_{23}|L_B} L_A^2 L_B^2 L_C \lambda^{L_A+L_B+L_c} . \qquad [107]$$

The integration of $h_2$, $J_{12}$ and $J_{23}$ can be performed first leading to a factor $\widetilde{P}_{L_A+L_B}^{J\neq 0}(J_{13}^{eff})$. We then find:

$$-2 a L_C \lambda^{L_C} \int_{-\infty}^{\infty} dJ_{13}^{eff} \int_{-\infty}^{\infty} dh_1 \int_{-\infty}^{\infty} dh_3 \, \widetilde{P}_{L_A+L_B}^{J\neq 0}(J_{13}^{eff}) \widehat{P}_3(h_1) \widehat{P}_3(h_3) \prod_{i=1,3} \Theta(|J_{13}^{eff}| - |h_i|) \delta(h_1 + h_3 \,\text{sign}(J_{13}^{eff}))$$

$$= -\frac{32 a^3 \left(\widehat{P}_3(0)\right)^3}{\rho^2} L_C \lambda^{L_A+L_B+L_C}. \qquad [108]$$

Note again the presence of the term $\widetilde{P}_{L_A+L_B}^{J\neq 0}(J_{13}^{eff})$ instead of $P_{L_A+L_B}^{J\neq 0}(J_{13}^{eff})$ due to our definition of the amputated diagrams. The other two contributions are the same, exchanging $L_C$ with $L_A$ and $L_B$. Adding this result to Eq. (106) we finally obtain:

$$\mathcal{C}_3^{lc}(\mathcal{G}_4^{amp}; L_A, L_B, L_C) = -\frac{32 a^3}{\rho^2} \left(\widehat{P}_3(0)\right)^3 \frac{L_A L_B + L_C L_B + L_A L_C}{L_A + L_B + L_C} \lambda^{L_A+L_B+L_C} . \qquad [109]$$

It is now easy to compute the full diagram with the three external legs: we have to multiply by

$$\frac{4 a \widehat{P}_1(0)}{\rho} \lambda^L \qquad [110]$$

for each line of length $L$. The result is:

$$\mathcal{C}_3^{lc}(\mathcal{G}_4; \vec{L}) = -\frac{2048 a^6}{\rho^5} \left(\widehat{P}_1(0)\right)^3 \left(\widehat{P}_3(0)\right)^3 \frac{L_A L_B + L_C L_B + L_A L_C}{L_A + L_B + Lc} \lambda^{L_1+L_2+L_3+L_A+L_B+L_C} . \qquad [111]$$

Now we have all the ingredients to compute the susceptibilities as shown in the main text. In a following section we will discuss additional and irrelevant diagrams.

**C. Susceptibilities on the lattice.** Here we want to make use of the results obtained for the three observables on the given diagrams to compute the susceptibilities shown in the main text, in Eqs. (28), (29), and (30). The first step is to write the Fourier transform of Eq. (88), Eq. (89) and Eq. (91). In order to precisely consider all the microscopic factors, such as the lattice spacing $a_l$, we define here the convention for the Fourier transform in the $D$ dimensional hypercubic grid, denoted with $a_l \times \mathbb{Z}^D$:

$$\widehat{f}(k) = a_l^D \sum_{x \in a_l \times \mathbb{Z}^D} f(x) e^{ikx} , \qquad f(x) = \int_{\left[-\frac{\pi}{a_l}, \frac{\pi}{a_l}\right]} \frac{d^D k}{(2\pi)^D} \widehat{f}(k) e^{-ikx} , \qquad [112]$$

that implies the expression of the Dirac delta function in the reciprocal space:

$$\left(\frac{2\pi}{a_l}\right)^D \delta^D(k) = \sum_{x \in a_l \mathbb{Z}^D} e^{ikx} . \qquad [113]$$

Now we define:

$$\widehat{C}_2(k_1, k_2) \equiv a_l^{2D} \sum_{x_1, x_2} e^{ik_1 x_1 + ik_2 x_2} C_2(x_1, x_2) \equiv (2\pi)^D \delta^D(k_1 + k_2) \widehat{C}_2(k_1),$$  [114]

$$\widehat{D}_2(k_1, k_2) \equiv a_l^{2D} \sum_{x_1, x_2} e^{ik_1 x_1 + ik_2 x_2} D_2(x_1, x_2) \equiv (2\pi)^D \delta^D(k_1 + k_2) \widehat{D}_2(k_1),$$  [115]

$$\widehat{C}_3(k_1, k_2, k_3) \equiv a_l^{3D} \sum_{x_1, x_2, x_3} e^{ik_1 x_1 + ik_2 x_2 + ik_3 x_3} C_3(x_1, x_2, x_3) \equiv (2\pi)^D \delta^D(k_1 + k_2 + k_3) \widehat{C}_3(k_1, k_2).$$  [116]

We will also use the following definition for the Fourier transform of the number of NBP:

$$\mathcal{N}_L(x_1, x_2) \equiv \int \frac{d^D k_1}{(2\pi)^D} \int \frac{d^D k_2}{(2\pi)^D} e^{-ik_1 x_1} e^{-ik_2 x_2} \widehat{\mathcal{N}}_L(k_1, k_2),$$  [117]

and, since the number of NBP only depends on the difference between the two points, we write

$$\widehat{\mathcal{N}}_L(k_1, k_2) = (2\pi)^D \delta^D(k_1 + k_2) \widehat{\mathcal{N}}_L(k_1).$$  [118]

Let us remark that we are interested in the critical point, that is dominated by the large-length behavior of the observables. For this reason we need the asymptotic expression of the NBP for small $k$ (37, 59):

$$\widehat{\mathcal{N}}_L(k) \approx 2D(2D-1)^{L-1} a_l^D e^{-k^2 a_l^2 L/(2D-2)}.$$  [119]

Now, starting from the Fourier transform of Eq. (88), Eq. (89) and Eq. (91), plugging Eq. (119) and the results of the observables on the given diagrams we obtain (up to negligible $1/M$ corrections):

$$\widehat{C}_2(k) = 4\rho\, a_l^D\, \frac{C B^2}{A} \left( \sum_{L=1}^{\infty} e^{-\left(\frac{a_l^2}{2D-2} k^2 + \tau\right) L} \right) \left( 1+ \right.$$
$$\left. - \frac{16 A a_l^D}{(2D-2)^{\frac{D}{2}}} \sum_{L=1}^{\infty} e^{-\left(\frac{a_l^2}{2D-2} k^2 + \tau\right) L} \sum_{L_A, L_B} \frac{L_A L_B}{L_A + L_B} \int \frac{d^D q}{(2\pi)^D} e^{-\left(\frac{a_l^2}{2D-2} q^2 + \tau\right) L_A - \left(\frac{a_l^2}{2D-2} (q-k)^2 + \tau\right) L_B} \right)$$  [120]

$$\widehat{D}_2(k) = -2\rho^2\, a_l^D\, \frac{C B^2}{A} \left( \sum_{L=1}^{\infty} L\, e^{-\left(\frac{a_l^2}{2D-2} k^2 + \tau\right) L} \right) \left( 1+ \right.$$
$$- \frac{32 A a_l^D}{(2D-2)^{\frac{D}{2}}} \left( \sum_{L=1}^{\infty} e^{-\left(\frac{a_l^2}{2D-2} k^2 + \tau\right) L} \right) \sum_{L_A, L_B} \frac{L_A L_B}{L_A + L_B} \int \frac{d^D q}{(2\pi)^D} e^{-\left(\frac{a_l^2}{2D-2} q^2 + \tau\right) L_A - \left(\frac{a_l^2}{2D-2} (q-k)^2 + \tau\right) L_B} +$$
$$\left. - \frac{16 A a_l^D}{(2D-2)^{\frac{D}{2}}} \frac{\left( \sum_{L=1}^{\infty} e^{-\left(\frac{a_l^2}{2D-2} k^2 + \tau\right) L} \right)^2}{\sum_{L=1}^{\infty} L\, e^{-\left(\frac{a_l^2}{2D-2} k^2 + \tau\right) L}} \sum_{L_A, L_B} L_A L_B \int \frac{d^D q}{(2\pi)^D} e^{-\left(\frac{a_l^2}{2D-2} q^2 + \tau\right) L_A - \left(\frac{a_l^2}{2D-2} (q-k)^2 + \tau\right) L_B} \right)$$  [121]

$$\widehat{C}_3(k_1, k_2) = 64\rho\, a_l^{2D}\, \frac{C B^3}{A} \prod_{i=1,2} \left( \sum_{L_i=1}^{\infty} e^{-\left(\frac{a_l^2}{2D-2} k_i^2 + \tau\right) L_i} \right) \left( \sum_{L_3=1}^{\infty} e^{-\left(\frac{a_l^2}{2D-2} (k_1+k_2)^2 + \tau\right) L_3} \right) \left( 1+ \right.$$
$$- \frac{16 A a_l^D}{(2D-2)^{\frac{D}{2}}} \sum_{L_A, L_B, L_C} \frac{L_A L_B + L_A L_C + L_B L_C}{L_A + L_B + L_C} \int \frac{d^D q}{(2\pi)^D} e^{-\left(\frac{a_l^2}{2D-2} q^2 + \tau\right) L_A - \left(\frac{a_l^2}{2D-2} (q-k_1)^2 + \tau\right) L_B - \left(\frac{a_l^2}{2D-2} (q+k_2)^2 + \tau\right) L_C}$$
$$\left. - \frac{16 A a_l^D}{(2D-2)^{\frac{D}{2}}} \left( \sum_{L=1}^{\infty} e^{-\left(\frac{a_l^2}{2D-2} k_2^2 + \tau\right) L} \right) \sum_{L_A, L_B} \frac{L_A L_B}{L_A + L_B} \int \frac{d^D q}{(2\pi)^D} e^{-\left(\frac{a_l^2}{2D-2} q^2 + \tau\right) L_A - \left(\frac{a_l^2}{2D-2} (q-k)^2 + \tau\right) L_B} \right)$$  [122]

where we used $\tau = -\ln((2D-1)\lambda)$ and we

$$A \equiv \frac{1}{M} \left( \frac{(2D)!}{(2D-3)!} \frac{\widehat{P}_3(0)}{\rho} \right)^2 \left( \frac{a}{(2D-1)2D} \right)^3 (2D-2)^{\frac{D}{2}},$$  [123]

$$B \equiv \frac{1}{M} \frac{(2D)!}{(2D-3)!} \frac{\widehat{P}_3(0)}{\rho} \frac{2D\, \widehat{P}_1(0)}{\rho} \left( \frac{a}{(2D-1)2D} \right)^2,$$  [124]

$$C \equiv (2D-2)^{\frac{D}{2}}.$$  [125]

Notice that, while $\rho$ and $a_l$ are dimensional quantities, with dimension of energy and length respectively, $\tau$ is dimensionless and it quantifies the distance from the Bethe lattice (mean-field) critical point, for this reason, it can be thought as the "bare" mass in field theory. Here we also notice that the lengths of the paths are dimensionless, they are only parameters for the number of NBP. The constants $A$, $B$ and $C$ are dimensionless too and the justification of their definitions is that, for a diagram with $N_{loop}$ loops and $V_e$ external vertices, a factor $A^{N_{loop}-1}$ and a factor $B^{V_e}$ respectively appear. The constant $C$ is defined only for brevity.

Another comment is on the perturbative nature of the expressions for $\widehat{C}_2$, $\widehat{D}_2$ and $\widehat{C}_3$. The $M$-layer construction provides a method to compute perturbative expansions for observables in the large $M$ limit so that the expansion parameter $1/M$ is reasonably small. For the sake of simplicity, from now on, we will drop the corrections, and the following computations are done neglecting higher orders in $1/M$ (or analogously $A$ or $B$, given that they both are $\mathcal{O}(1/M)$).

The quantities we are interested in are

$$\sum_x C_2(x,0) = \frac{1}{a_l^D} \widehat{C}_2(0) = \chi_2(\tau), \qquad [126]$$

$$\sum_{x,y} C_3(x,y,0) = \frac{1}{a_l^{2D}} \widehat{C}_3(0,0) = \chi_3(\tau), \qquad [127]$$

$$\sum_x D_2(x,0) = \frac{1}{a_l^D} \widehat{D}_2(0) = \chi_2^{dis}(\tau). \qquad [128]$$

Now we have all the ingredients to compute the susceptibilities to one-loop order, to do so we perform the integrals over the "loop momenta", denoted by $q$ in all the three expressions given in Eqs. (120), (121) and (122), then we may write the sums over the lengths as integrals

$$\prod_{i=1}^{I} \sum_{L_i=1}^{\infty} \to \prod_{i=1}^{I} \int_{1/\Lambda^2}^{\infty} dL_i, \qquad [129]$$

where, for later convenience, we preferred to explicitly write the small-length cut-off $\Lambda = 1$. Then we can scale the lengths

$$\widetilde{L}_i \equiv L_i \tau \qquad [130]$$

and, plugging Eqs. (120), (121), (122) inside the definitions of the susceptibilities, we have

$$\chi_2(\tau) = 4\frac{\rho}{\tau} C \frac{B^2}{A} \left(1 - \frac{16 A \tau^{\frac{D}{2}-4}}{(4\pi)^{D/2}} I_1(\tau/\Lambda^2)\right), \qquad [131]$$

$$\chi_3(\tau) = 64 \frac{\rho}{\tau^3} C \frac{B^3}{A} \left(1 - \frac{32 A \tau^{\frac{D}{2}-4}}{(4\pi)^{D/2}} I_3(\tau/\Lambda^2) - \frac{48 A \tau^{\frac{D}{2}-4}}{(4\pi)^{D/2}} I_1(\tau/\Lambda^2)\right), \qquad [132]$$

$$\chi_2^{dis}(\tau) = -2 \frac{\rho^2}{\tau^2} C \frac{B^2}{A} \left(1 - \frac{32 A \tau^{\frac{D}{2}-4}}{(4\pi)^{D/2}} \left(I_1(\tau/\Lambda^2) + I_2(\tau/\Lambda^2)\right)\right), \qquad [133]$$

where we defined the following integrals

$$I_1(\tau/\Lambda^2) \equiv \int_{\tau/\Lambda^2}^{\infty} d\widetilde{L}_a \, d\widetilde{L}_b \, \frac{\widetilde{L}_a \widetilde{L}_b \, e^{-(\widetilde{L}_a+\widetilde{L}_b)}}{(\widetilde{L}_a + \widetilde{L}_b)^{D/2+1}}, \qquad [134]$$

$$I_2(\tau/\Lambda^2) \equiv \int_{\tau/\Lambda^2}^{\infty} d\widetilde{L}_a \, d\widetilde{L}_b \, \frac{\widetilde{L}_a^2 \widetilde{L}_b \, e^{-(\widetilde{L}_a+\widetilde{L}_b)}}{(\widetilde{L}_a + \widetilde{L}_b)^{D/2+1}} = \frac{1}{2} \int_{\tau/\Lambda^2}^{\infty} d\widetilde{L}_a \, d\widetilde{L}_b \, \frac{\widetilde{L}_a \widetilde{L}_b \, e^{-(\widetilde{L}_a+\widetilde{L}_b)}}{(\widetilde{L}_a + \widetilde{L}_b)^{D/2}}, \qquad [135]$$

and

$$I_3(\tau/\Lambda^2) \equiv \int_{\tau/\Lambda^2}^{\infty} d\widetilde{L}_a \, d\widetilde{L}_b \, d\widetilde{L}_c \, \frac{\widetilde{L}_a \widetilde{L}_b + \widetilde{L}_c \widetilde{L}_b + \widetilde{L}_a \widetilde{L}_c}{(\widetilde{L}_a + \widetilde{L}_b + \widetilde{L}_c)^{D/2+1}} e^{-(\widetilde{L}_a+\widetilde{L}_b+\widetilde{L}_c)}. \qquad [136]$$

Note that $I_1(x)$ diverges for $x \to 0$ (ultraviolet regime) in $D = 8$, thus further complicating the expansion at small values of $\tau$, while $I_2(x)$ and $I_3(x)$ are regular. These typical divergences can be cured by writing the susceptibilities as functions of the physical mass $m$, that is defined as the reciprocal of the correlation length as follows. The correlation length is defined from the connected correlation through the following formula:

$$\xi^2 = \left(\frac{d}{dk^2} \widehat{C}_2^{-1}(k)\right)\bigg|_{k=0} \widehat{C}_2(0), \qquad [137]$$

where, from Eq. (120), scaling the momenta

$$\widetilde{k} \equiv k \frac{a_l}{\sqrt{\tau(2D-2)}}, \qquad [138]$$

we can write

$$\widehat{C}_2(k) = a_l^D \frac{\rho}{\tau} C \frac{B^2}{A} \left(4 \frac{1}{\widetilde{k}^2+1} - \frac{128}{2} A \tau^{\frac{D}{2}-4} \left(\frac{1}{\widetilde{k}^2+1}\right)^2 \int \frac{d^D \widetilde{q}}{(2\pi)^D} \int_{\tau/\Lambda^2}^{\infty} d\widetilde{L}_a \, d\widetilde{L}_b \, \frac{\widetilde{L}_a \widetilde{L}_b}{\widetilde{L}_a + \widetilde{L}_b} e^{-\widetilde{q}^2 \widetilde{L}_a - (\widetilde{q}+\widetilde{k})^2 \widetilde{L}_b - (\widetilde{L}_a+\widetilde{L}_b)}\right). \qquad [139]$$

Then we need

$$\widehat{C}_2(0) = a_l^D \frac{\rho}{\tau} C \frac{B^2}{A} \left(4 - \frac{128}{2} A \tau^{\frac{D}{2}-4} \frac{1}{(4\pi)^{D/2}} \int_{\tau/\Lambda^2}^{\infty} d\widetilde{L}_a \, d\widetilde{L}_b \, \frac{\widetilde{L}_a \widetilde{L}_b}{(\widetilde{L}_a + \widetilde{L}_b)^{D/2+1}} e^{-(\widetilde{L}_a+\widetilde{L}_b)}\right), \qquad [140]$$

$$\frac{1}{\widehat{C}_2(k)} = a_l^{-D} \frac{\tau}{\rho} \frac{A}{C B^2} \frac{\widetilde{k}^2+1}{4} \left(1 - 16 A \tau^{\frac{D}{2}-4} \frac{1}{\widetilde{k}^2+1} \int \frac{d^D \widetilde{q}}{(2\pi)^D} \int_{\tau/\Lambda^2}^{\infty} d\widetilde{L}_a d\widetilde{L}_b \frac{\widetilde{L}_a \widetilde{L}_b}{\widetilde{L}_a + \widetilde{L}_b} e^{-\widetilde{q}^2 \widetilde{L}_a - (\widetilde{q}+\widetilde{k})^2 \widetilde{L}_b - (\widetilde{L}_a + \widetilde{L}_b)}\right)^{-1}$$

$$= a_l^{-D} \frac{\tau}{\rho} \frac{A}{C B^2} \frac{\widetilde{k}^2+1}{4} \left(1 + 16 A \tau^{\frac{D}{2}-4} \frac{1}{\widetilde{k}^2+1} \int \frac{d^D \widetilde{q}}{(2\pi)^D} \int_{\tau/\Lambda^2}^{\infty} d\widetilde{L}_a d\widetilde{L}_b \frac{\widetilde{L}_a \widetilde{L}_b}{\widetilde{L}_a + \widetilde{L}_b} e^{-\widetilde{q}^2 \widetilde{L}_a - (\widetilde{q}+\widetilde{k})^2 \widetilde{L}_b - (\widetilde{L}_a + \widetilde{L}_b)}\right)$$

$$= a_l^{-D} \frac{\tau}{\rho} \frac{A}{C B^2} \frac{1}{4} \left(\widetilde{k}^2 + 1 + \frac{16 A \tau^{\frac{D}{2}-4}}{(4\pi)^{D/2}} \int_{\tau/\Lambda^2}^{\infty} d\widetilde{L}_a d\widetilde{L}_b \frac{\widetilde{L}_a \widetilde{L}_b}{(\widetilde{L}_a + \widetilde{L}_b)^{D/2+1}} \exp\left(-\widetilde{k}^2 \frac{\widetilde{L}_a \widetilde{L}_b}{\widetilde{L}_a + \widetilde{L}_b} - (\widetilde{L}_a + \widetilde{L}_b)\right)\right), \quad [141]$$

and also

$$\frac{\partial}{\partial \widetilde{k}^2} \widehat{C}_2(k)^{-1}\bigg|_{\widetilde{k}^2=0} = a_l^{-D} \frac{\tau}{\rho} \frac{A}{C B^2} \frac{1}{4} \left(1 - \frac{16 A \tau^{\frac{D}{2}-4}}{(4\pi)^{D/2}} \int_{\tau/\Lambda^2}^{\infty} d\widetilde{L}_a d\widetilde{L}_b \frac{\widetilde{L}_a^2 \widetilde{L}_b^2}{(\widetilde{L}_a + \widetilde{L}_b)^{D/2+2}} \exp\left(-(\widetilde{L}_a + \widetilde{L}_b)\right)\right), \quad [142]$$

to compute

$$\xi^2 = \left(\frac{\partial}{\partial k^2} \widehat{C}_2^{-1}(k)\right)\bigg|_{k=0} \widehat{C}_2(0) = \widehat{C}_2(0) a_l^2 C^{-2/D} \frac{1}{\tau} \frac{\tau}{\rho} \frac{A}{C B^2} \frac{1}{4} \left(1 - \frac{16 A \tau^{\frac{D}{2}-4}}{(4\pi)^{D/2}} \int_{\tau/\Lambda^2}^{\infty} d\widetilde{L}_a d\widetilde{L}_b \frac{\widetilde{L}_a^2 \widetilde{L}_b^2}{(\widetilde{L}_a + \widetilde{L}_b)^{D/2+2}} e^{-(\widetilde{L}_a + \widetilde{L}_b)}\right)$$

$$= a_l^2 C^{-2/D} \frac{1}{\tau} \left(1 - \frac{16 A \tau^{\frac{D}{2}-4}}{(4\pi)^{D/2}} \left(I_4(\tau/\Lambda^2) + I_1(\tau/\Lambda^2)\right)\right), \quad [143]$$

where

$$I_4(\tau/\Lambda^2) \equiv \int_{\tau/\Lambda^2}^{\infty} d\widetilde{L}_a d\widetilde{L}_b \frac{\widetilde{L}_a^2 \widetilde{L}_b^2 e^{-(\widetilde{L}_a + \widetilde{L}_b)}}{(\widetilde{L}_a + \widetilde{L}_b)^{D/2+2}}. \quad [144]$$

Note that $I_4(x)$ is finite for $x \to 0$ in $D=8$. Analogously we can write the last equation in terms of the physical distance from the critical point, which we call $m$

$$m^2 = \frac{1}{\xi^2} = \frac{C^{2/D}}{a_l^2} \tau \left(1 + \frac{16 A \tau^{\frac{D}{2}-4}}{(4\pi)^{D/2}} \left(I_4(\tau/\Lambda^2) + I_1(\tau/\Lambda^2)\right)\right). \quad [145]$$

From the last equation we can write $\tau$ as a function of the mass $m$ (neglecting higher orders in $A$):

$$\tau = a_l^2 C^{-2/D} m^2 \left(1 - \frac{1}{(4\pi)^{D/2}} \frac{u}{2} \left(I_4(m^2/\Lambda^2) + I_1(m^2/\Lambda^2)\right)\right), \quad [146]$$

where we defined the dimensionless coupling constant $u$

$$u \equiv g\, m^{D-8} \quad [147]$$

and

$$g \equiv 32 A \, (a_l C^{-1/D})^{D-8}. \quad [148]$$

Given the last definition and using Eq. (146) we can rewrite the three observables as functions of $m^2$, as in the main text

$$\chi_2(m^2) = 4\rho\, C^{1+\frac{2}{D}} a_l^{-2} m^{-2} \frac{B^2}{A} \left(1 + \frac{1}{(4\pi)^{D/2}} \frac{u}{2} I_4(m^2/\Lambda^2)\right), \quad [149]$$

$$\chi_3(m^2) = 64\rho\, C^{1+\frac{6}{D}} a_l^{-6} m^{-6} \frac{B^3}{A} \left(1 + \frac{u}{(4\pi)^{D/2}} \left(\frac{3}{2} I_4(m^2/\Lambda^2) - I_3(m^2/\Lambda^2)\right)\right) \quad [150]$$

and

$$\chi_2^{dis}(m^2) = -2\rho^2\, C^{1+\frac{4}{D}} a_l^{-4} m^{-4} \frac{B^2}{A} \left(1 + \frac{u}{(4\pi)^{D/2}} \left(I_4(m^2/\Lambda^2) - I_2(m^2/\Lambda^2)\right)\right). \quad [151]$$

Note that upon switching from the bare parameter $\tau$ to the physical correlation length $1/m$ the integral $I_1(\tau/\Lambda^2)$, that is ultraviolet divergent in $D=8$, is no longer present and the expressions are free of ultraviolet divergences, providing the possibility to safely take the limit $m \to 0$.

**C.1. Other irrelevant diagrams.** In this section we argue about the irrelevance of some diagrams that in principle we should consider to compute the observables of interest.

The first kind of diagrams we neglected is the "tadpole-type" ones, depicted in Fig. S1. In principle such topologies should be considered not only for two-point functions, but also to "dress" the external or internal lines of a generic diagram for a generic $n$-point function, for instance, the diagram $\mathcal{G}_3$ for the three-point function we computed. The justification for neglecting this kind of topologies is simple and makes use of the property of the *Ansatz* distribution of the line, Eq. (8) in the main text. Let us consider two lines, of the kind of $\mathcal{G}_1$, joined by a central common spin $\sigma_0$ and with $\sigma_1$ and $\sigma_2$ at the extremities of the chain. Now, instead of simply joining them as done before to impose the self-consistency of the *Ansatz*, let us imagine adding a line starting and ending in $\sigma_0$, thus forming a four-degree vertex and a loop. In order to compute the distribution of the effective fields on $\sigma_1$ and $\sigma_2$ and the effective coupling between the two, we have to sum over the configurations of $\sigma_0$, which depends on the parameters of each of the three lines. In the end we will integrate over these parameters following some rules which are similar to the ones in Tab. S1. However the coupling of the loop line $J_{00}$ is ineffective since its contribution to the effective Hamiltonian, $J_{00}\sigma_0^2$, is independent of $\sigma_0$, given the fact that $\sigma_0^2=1$. This means that, integrating on $J_{00}$, without any constraint, the *Ansatz* distribution of the loop line, only the trivial part will survive

$$\int dJ_{00} \left(\delta(J_{00}) P_B(u_1) P_B(u_2) - 2a\, L\, \lambda^L \delta(J) g(u_1) g(u_2) + P_L^{J \neq 0}(J_{00}) g(u_1) g(u_2)\right) = P_B(u_1) P_B(u_2), \quad [152]$$

since, due to the above-mentioned property, the integral over the coupling of the non-trivial zero-coupling part is equal and opposite in sign of the integral of the non-zero part. We understand that to our order in the large-length limit, closing a line on a site doesn't change

the *Ansatz* distribution (a similar argument can be applied to the three-degree vertex tadpole-type diagram). Since, these tadpoles are one-loop diagrams we have to be careful, in the $M$-layer framework, and apply the line-connected definition, which amounts to subtracting the simple line without the loop. The result is then zero, given that the distribution of the loop diagram is the same as the simple line. For this reason, every complicated diagram, dressed with such topologies gives zero contribution to the observable for the spin glass model in a field at zero temperature.

Another kind of diagram we completely neglected is $\mathcal{G}'''$ in Fig. S2. This diagram can be seen as a special case of $\mathcal{G}_5$ where the internal line connecting the loop to the two other external lines has zero length. However, we remember that the minimum length for a line is always 1, in order to correctly count all the combinatorial factors originating in each vertex. Indeed, $\mathcal{G}_5$ has three three-degree vertices, while contracting the internal line two three-degree vertices become a single four-degree vertex. For this reason, in principle we should separately compute the contributions of all diagrams originated from the contraction of a line of a generic diagram. Luckily we can argue that every time we contract a line the resulting contribution misses a propagator ($(k^2 + \tau)^{-1}$ for $k^2 \sim 0$), thus it will come with a multiplicative factor $\tau$ with respect to the "non-contracted" contribution. When we compute the critical exponents we want to perform the critical limit $\tau \to 0$ (or equivalently $m^2 \to 0$), which is why these "contracted" diagrams are not relevant at the critical limit to our order in the large-length limit, because they will give a subdominant contribution.

**C.2. Generic n-point connected function at one loop.** Even if we computed all relevant diagrams we want to extend the idea of $P^{NB}_{\{\bar{L}\}}$ in order to compute the contribution to a generic $n$-point connected function on diagrams of the kind depicted in Fig. S4.

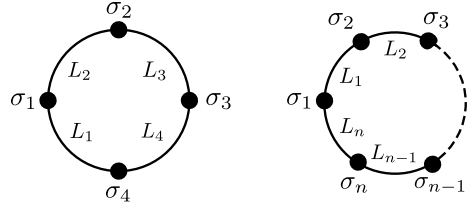

**Fig. S4.** Left: four-point diagram computed in this section. Right: relevant diagram for the amputated $n$-point function: $\mathcal{G}^{amp}_n$.

We will consider the case $n = 4$ and a simple generalization leads to the generic formula for the $n$-point function. As for the three-point function, we start by cutting one line, for instance, the one with length $L_4$ as in Fig. S5, we compute the probability that the four spins belong to the same cluster in this simple case and then we reintroduce the line in a second moment.

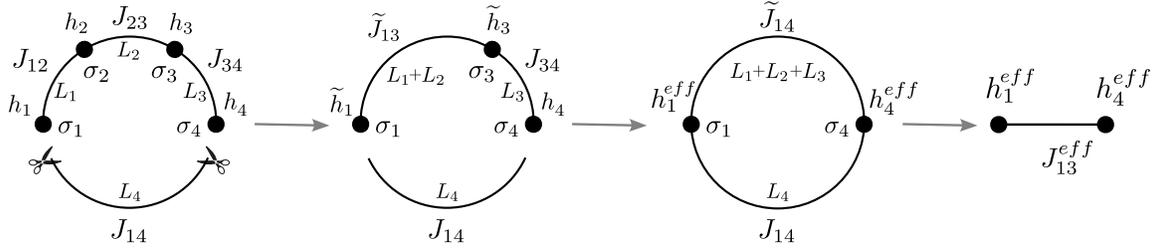

**Fig. S5.** Representation of the steps of the procedure to compute the corresponding $P^{NB}_{L_1,\ldots,L_{n-1}}$ distribution on the specific case $n = 4$.

Again, we have to take into account two possibilities when re-introducing $L_4$: i) the coupling $J_{14}$ is such that the ground state without the line is not altered and ii) the coupling $J_{14}$ is such that the ground state is altered.

As for the three-point function let us start considering the case where all the couplings are different from zero. The first case is simple: we can sum over the configurations of $\sigma_2$ and $\sigma_3$, putting one external field drawn from $g(u)$ in each vertex, to obtain

$$\left(\frac{4\, a\, \widehat{P}_3(0)}{\rho}\right)^2 P^{J \neq 0}_{L_1+L_2+L_3}(\widetilde{J}_{14}) \widehat{P}_3(h_1^{eff}) \widehat{P}_3(h_4^{eff}) \qquad [153]$$

and then add the line of length $L_4$ with the usual condition for the resulting effective coupling and fields: $J^{eff}_{14} = \widetilde{J}_{14} + J_{14}$, $h_1^{eff}$ and $h_4^{eff}$. As for the three-point function case we notice that summing over the configurations of $\sigma_2$ (before) and $\sigma_3$ (after), at our level of approximation in the large length limit, is ineffective on the values of the effective fields: $h_1^{eff} \simeq h_1$, $\widetilde{h}_3 \simeq h_3$ and $h_4^{eff} \simeq h_4$, see Fig. S5. The result for this first contribution is:

$$2 \int_0^\infty dJ^{eff}_{14} \int_0^\infty d\widetilde{J}_{14} \int_{-J_{14}}^\infty dJ_{14} \int_{-\infty}^\infty dh_1 \int_{-\infty}^\infty dh_4 \, P^{J \neq 0}_{L_1+L_2+L_3}(\widetilde{J}_{14}) \, P^{J \neq 0}_{L_3}(J_{13}) \, \widehat{P}_3(h_1) \, \widehat{P}_3(h_4) \delta(J^{eff}_{14} - (\widetilde{J}_{14} + J_{14})) \times$$

$$\times \left(\frac{4\, a\, \widehat{P}_3(0)}{\rho}\right)^2 \prod_{i=1,4} \Theta(|J^{eff}_{14}| - |h_i|) \, \delta(|h_1 + h_4 \operatorname{sign}(J^{eff}_{14})|) =$$

$$= \frac{64 \, a^4 \left(\widehat{P}_3(0)\right)^4}{\rho^3} \frac{L_1^2 + L_2^2 + L_3^3 + 2L_4^3 + 2L_1 L_2 + 2L_2 L_3 + 2L_1 L_3 + 2L_1 L_4 + 2L_2 L_4 + 2L_3 L_4}{L_1 + L_2 + L_3 + L_4} \lambda^{L_1+L_2+L_3+L_4}, \qquad [154]$$

where we considered explicitly the case of positive couplings, the other case is the same by symmetry and gives the factor 2 in front. The same symmetry arguments are used for the next contribution.

The contribution of the second case is the non-trivial one. Again we start by removing $L_4$ and then take into account the scenario in which all the four spins belong to the same cluster even if the coupling $J_{14}$, when re-introduced, changes the ground state configuration of $\sigma_1$ and $\sigma_4$ from the one where $J_{14} = 0$. To do so we want to first join the lines $L_1$ and $L_2$, resulting in the effective coupling $\widetilde{J}_{13}$. Next we should add $L_3$ with the condition that $|h_3| < |\widetilde{J}_{13}| - |J_{34}|$ if $|\widetilde{J}_{13}| > |J_{34}|$ or $|h_3| < |J_{34}| - |\widetilde{J}_{13}|$ $|\widetilde{J}_{13}| < |J_{34}|$ otherwise. We notice that $\sigma_3$ will follow $\sigma_1$ if $|\widetilde{J}_{13}| > |J_{34}|$, otherwise it will follow $\sigma_2$. This means that in the first situation, it is sufficient to consider $P^{J\neq 0}_{L_1+L_2}(\widetilde{J}_{13})$, since the necessary condition for $\sigma_3$ to follow $\sigma_1$ is that $\sigma_2$ follows too, while in the second situation, since $\sigma_3$ follows $\sigma_4$, we should be sure that $\sigma_2$ is "non-blocked", thus we have to use $P^{NB}_{L_1,L_2}(\widetilde{J}_{13})$. With this argument we can compute $P^{NB}_{L_1,L_2,L_3}(\widetilde{J}_{14})$, the result of the sum of the two following contributions

$$\frac{4a\left(\widehat{P}_3(0)\right)^2}{\rho}\int_{J_{34}}^{\infty} d\widetilde{J}_{13}\int_{-\infty}^{\infty} dJ_{34}\, P^{J\neq 0}_{L_1+L_2}(\widetilde{J}_{13})\, P^{J\neq 0}_{L_3}(J_{34})\delta(\widetilde{J}_{14} - J_{34})\, 2\,(\widetilde{J}_{13} - J_{34}) +$$

$$+\frac{4a\left(\widehat{P}_3(0)\right)^2}{\rho}\int_{\widetilde{J}_{13}}^{\infty} dJ_{34}\int_{-\infty}^{\infty} d\widetilde{J}_{13}\, P^{NB}_{L_1,L_2}(\widetilde{J}_{13})\, P^{J\neq 0}_{L_3}(J_{34})\delta(\widetilde{J}_{14} - \widetilde{J}_{13})\, 2\,(J_{34} - \widetilde{J}_{13}),\quad [155]$$

which is a sort of generalization of Eq. (102). The result is

$$P^{NB}_{L_1,L_2,L_3}(\widetilde{J}_{14}) = \frac{16\,a^3\left(\widehat{P}_3(0)\right)^2}{\rho}\,(L_1^2 + L_2^2 + L_3^2)\,\lambda^{L_1+L_2+L_3}\,e^{-\rho|\widetilde{J}_{14}|(L_1+L_2+L_3)}.\quad [156]$$

Then we can compute the contribution of the second case:

$$2\int_{-\infty}^{0} dJ^{eff}_{14}\int_{0}^{\infty} d\widetilde{J}_{14}\int_{-\infty}^{-\widetilde{J}_{14}} dJ_{14}\int_{-\infty}^{\infty} dh_1\int_{-\infty}^{\infty} dh_4\, P^{NB}_{L_1,L_2,L_3}(\widetilde{J}_{14})\, P^{J\neq 0}_{L_4}(J_{14})\widehat{P}_2(h_1)\widehat{P}_2(h_4)\times$$

$$\times\prod_{i=1,4}\Theta(|J^{eff}_{14}| - |h_i|)\,\delta(|h_1 + h_4\,\text{sign}(J^{eff}_{14})|) = \frac{64\,a^4\left(\widehat{P}_3(0)\right)^4}{\rho^3}\frac{L_1^2+L_2^2+L_3^2}{L_1+L_2+L_3+L_4}\lambda^{L_1+L_2+L_3+L_4}.\quad [157]$$

The last contribution is given by the case in which one of the four lines has zero coupling:

$$-\frac{128\,a^4\left(\widehat{P}_3(0)\right)^4}{\rho^3}\,(L_1+L_2+L_3+L_4).\quad [158]$$

All the terms together give this contribution to the four-point function

$$\mathcal{C}^{lc}_4(\mathcal{G}^{amp}_6; L_A, L_B, L_C, L_D) = -\frac{128\,a^4}{\rho^3}\left(\widehat{P}_3(0)\right)^4\frac{L_AL_B + L_CL_B + L_AL_C + L_AL_D + L_BL_D + L_CL_D}{L_A+L_B+L_C+L_D}\lambda^{L_A+L_B+L_C+L_D}.\quad [159]$$

From this result is not difficult to express the generic form of the $P^{NB}$ for a generic line of length $L_1 + L_2 + \cdots + L_{n-1}$ generalizing the expression of Eq. (155)

$$P^{NB}_{L_1,\ldots,L_{n-1}}(\widetilde{J}_{1n}) = a\,\rho\left(\frac{4\,a\,\widehat{P}_3(0)}{\rho}\right)^{n-2}\sum_{i=1}^{n-1}L_i^2\prod_{i=1}^{n-1}\lambda^{L_i}\,e^{-\rho|\widetilde{J}_{1n}|L_i},\quad [160]$$

from which we are able to write the contribution of the diagram depicted in Fig. S4 for a generic $n$-point function:

$$\mathcal{C}^{lc}_n(\mathcal{G}^{amp}_n; \{L_i\}_{i=1,\ldots,n}) = -\frac{\rho}{2}\left(\frac{4\,a\,\widehat{P}_3(0)}{\rho}\right)^n\frac{\sum_{i<j}^{1,\ldots,n}L_i\,L_j}{\sum_{i=1}^{n}L_i}\prod_{i=1}^{n}\lambda^{L_i}.\quad [161]$$

### 3. Numerical tests

In this section, we report some numerical tests we did to confirm the perturbative expressions we obtained for the observables on the given diagrams, in Eq. (79), Eq. (77), Eq. (95), Eq. (96), Eq. (85), Eq. (97), and Eq. (111). To do so we used the numerical technique described in the SI of (40). We summarize here the main points. Notice that the following numerical results are obtained for Gaussian couplings $J \sim \mathcal{N}(0, 1/(z-1))$ and fixed external field $H$ on each site, where $z$ is the connectivity of the graph. Due to universality, we believe that the bimodal distributed coupling and Gaussian external field case would give the same results.

Given the rules for the evolution in Tab. S1, the idea is to start from a population of $N_T = 10^7$ triplets of two fields and one coupling characterized by length $L = 1$, *i.e.* distributed as $P_{L=1}(u_0, u_1, J_1) = \delta(u_0)\delta(u_1)\mathcal{N}(0, 1/(z-1))$. Joining one triplet of length $L - 1$ with one of the kind $(0, 0, J)$, one construct iteratively a new triplet of length $L$. We can easily see, from the evolution rules in Tab. S1, that at each step a fraction of the population that satisfies $|h| > |J_{L-1}| + |J|$, produces a new one with zero coupling, $J_L = 0$. Given that the effective coupling of two infinitely distant sites is zero, the size of the population with $J_L \neq 0$ shrinks to zero exponentially fast with $L$. Moreover, for the spin glass in a field the triplets with zero coupling, $(u_0, u_L, 0)$, can be of two different kinds. Indeed, when one joins a triplet of the type $(u_0, u_L, 0)$, with $u_0$ and $u_L$ that are correlated, with one of the type $(0, 0, J)$, looking at the case $|h| > |J|$ of Tab. S1, the correlation between the new fields $u_0$ and $u_{L+1}$ is zero because of the random sign of $J$. In total, three different signals arise: one with non-zero coupling, one with zero coupling but correlated fields and one with zero coupling and uncorrelated fields. Thus, to amplify these three distinct signals, we evolve three populations of the same size $N_T$ for each $L$ with an enrichment procedure, alongside with their probabilities $p_L$ and $b_L$ that are the weights of $\mathcal{A}_L$ and $\mathcal{B}_L$ respectively: $\mathcal{A}_L$ stores the triplets with $J_L \neq 0$, $\mathcal{B}_L$ keeps correlated fields $(u_0, u_L)$ but $J_L = 0$ and $\mathcal{C}_L$ that keeps uncorrelated pairs $(u_0, u_L)$. Joining a triplet of length $L - 1$, denoted as $T_{L-1}$, to one of length 1 of the type $(0, 0, J)$, to form a new longer triplet, $T_L$, we have different cases:

- $T_{L-1} \in \mathcal{A}_{L-1}$ and $|h| < |J_{L-1}| + |J|$ → $T_L \in \mathcal{A}_L$ ,

- $T_{L-1} \in \mathcal{A}_{L-1}$ and $|h| > |J_{L-1}| + |J| \quad \rightarrow \quad T_L \in \mathcal{B}_L$ ,

- $T_{L-1} \in \mathcal{B}_{L-1}$ and $|h| < |J| \quad \rightarrow \quad T_L \in \mathcal{B}_L$ ,

in all other cases $T_L \in \mathcal{C}_L$. The associated initial probabilities are $p_1 = 1$, $b_1 = 0$. To compute the Bethe distribution of cavity fields $P_B(u)$ we used the corresponding cavity equations (at zero temperature) with 1000 iterations for a population of $10^6$ fields.

Once the population of the triplet for each line is obtained, we can add up the lines summing over the values of the internal spins in order to generate the distributions for the diagrams we need to compute and estimate the values of the observables. Particularly easy is the case of the disconnected function (line-connected), since it corresponds to the zero-coupling part of the distribution (once the trivial $P_B(u_1) P_B(u_2) \delta(J)$ part is subtracted). Regarding the connected functions, we have some additional work to do. Once the distribution of the effective parameters is obtained, from the population of triplets, we have to average the connected function. To do so, given the triplet, we count 1 every time the external spins belong to the same cluster. The result is averaged over the distribution of the triplets. We then repeated this computation $N_{data}$ times for each observable and each set of lengths of the given diagram, each time extracting independently a new distribution of the triplets and we computed the mean value and its associated statistical error. The result corresponds to the connected functions we computed analytically.

As we anticipated for the *Ansatz* distribution of the triplet, each parameter can be computed or at least numerically estimated. Notice that here and for the following results we set the connectivity of the Bethe lattice $z = 3$ and the associated critical field to $H = 0.358$. In Fig. S7 we show the results for the estimate of the parameter $a$ of the *Ansatz*, in Eq. (46). To do so we used the two most numerically precise results, that are for $\mathcal{C}_2^{lc}(\mathcal{G}_1; L)$ and $\mathcal{D}_2^{lc}(\mathcal{G}_1; L)$. With the former we estimated the value of $\lambda$ as shown in Fig. S6 fitting the data with free parameter $\alpha$, which is the multiplicative factor of the term $\lambda^L$, see Eq. (79). Then we used it to fit $\mathcal{D}_2^{lc}(\mathcal{G}_1; L)$ and extract the value of $a$. We see that the extrapolated values are compatible with the values $\lambda = 1/2$ and $\alpha = 1$. $\lambda = 1/2$ is expected exactly at the

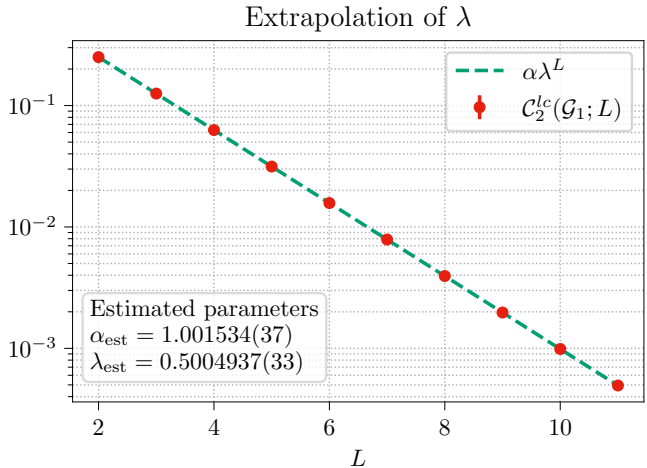
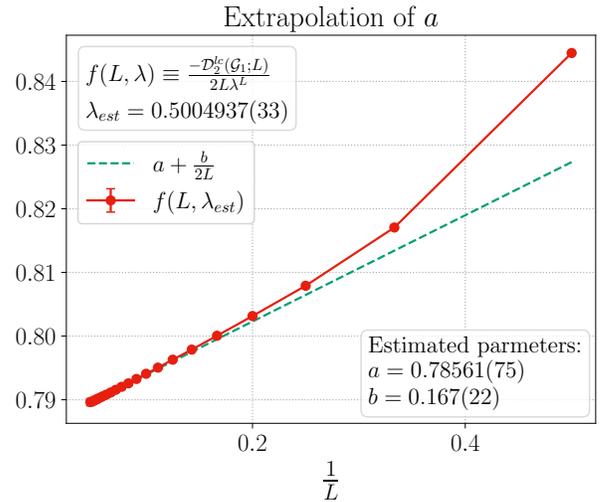

**Fig. S6.** Numerical extrapolation of the control parameter $\lambda$. Points are the data while the dashed line is the fitting function with parameters $\lambda_{est}$ and $\alpha_{est}$.

**Fig. S7.** Numerical extrapolation of the Ansatz parameter $a$. Points come from the numerical evaluation of the bare two-points disconnected correlation, scaled as $\frac{-\mathcal{D}_2^{lc}(\mathcal{G}_1; L)}{2L\lambda^L}$, with the value of $\lambda$ obtained from Fig. S6. The dashed line is the fitting function that allows us to extrapolate the Ansatz parameter $a$.

critical point and the observed deviation is due to a not fully precise value of the critical field $H = 0.358$. $\alpha = 1$ is expected because the parameter $\alpha$ can be expressed as the ratio between $g(u)$ and $P_B(u)$ and we see clear numerical evidence that the function $g(u) = P_B(u)$ in the Gaussian couplings case at the critical point. The deviation of the fitted value of $\alpha$ from 1 can be due to the finite size of the population used.

In Fig. S8 we re-obtained the numerical results of ref. (40) for the two-point connected function. In particular we performed a joint fit of the parameters, whose results are shown in the table on the right and, for large lengths, are in good agreement with the analytical computation. Notice that the amputated two-point connected function on $\mathcal{G}_2$ depends on two lengths, $L_A$ and $L_B$. To check the result in a single plot we numerically computed the observable for two different combinations of lengths: $L_A = L_B$ and $L_B = 2L_A$.

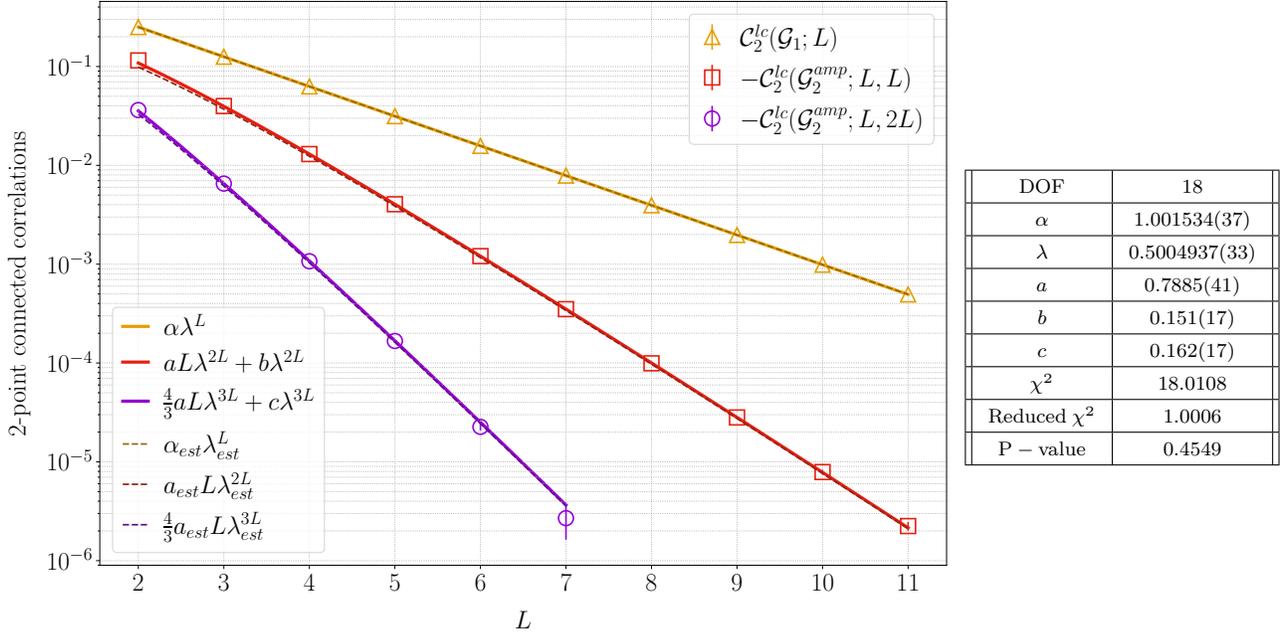

**Fig. S8.** Left: Points correspond to the numerical evaluation of the two-point connected correlation function on the different diagrams of Fig. S1 as a function of the length $L$; full lines correspond to the associated joint fit (that also includes next-to-leading order term w.r.t. $L$); dashed lines correspond to the analytical prediction, with the values $\alpha_{est}$, $\lambda_{est}$ and $a_{est}$ extracted from Fig. S7. Right: the evaluated corresponding fit parameters.

In Fig. S8 we report the numerical evaluation of the two-point connected correlation function. Data have been fitted with the following functions and parameters:

$$\mathcal{C}_2^{lc}(\mathcal{G}_1; L) = \alpha\lambda^L \quad \text{for } L \in [2, 11], \quad N_{data} = 200\,; \qquad [162]$$

$$-\mathcal{C}_2^{lc}(\mathcal{G}_2^{amp}; L, L) = aL\lambda^{2L} + b\lambda^{2L} \quad \text{for } L \in [4, 11], \quad N_{data} = 200\,; \qquad [163]$$

$$-\mathcal{C}_2^{lc}(\mathcal{G}_2^{amp}; L, 2L) = \frac{4}{3}aL\lambda^{3L} + c\lambda^{3L} \quad \text{for } L \in [3, 7], \quad N_{data} = 400\,. \qquad [164]$$

In Fig. S9 we show the results for the three-point function. In particular we plot the bare contribution together with four different combinations of lengths for the amputated diagram $\mathcal{G}_4^{amp}$. The results of the associated joint fit are collected in the table on the right. Also in this new case we confirm the good agreement between the numerical and the analytical results for the three-point connected function on the given diagrams.

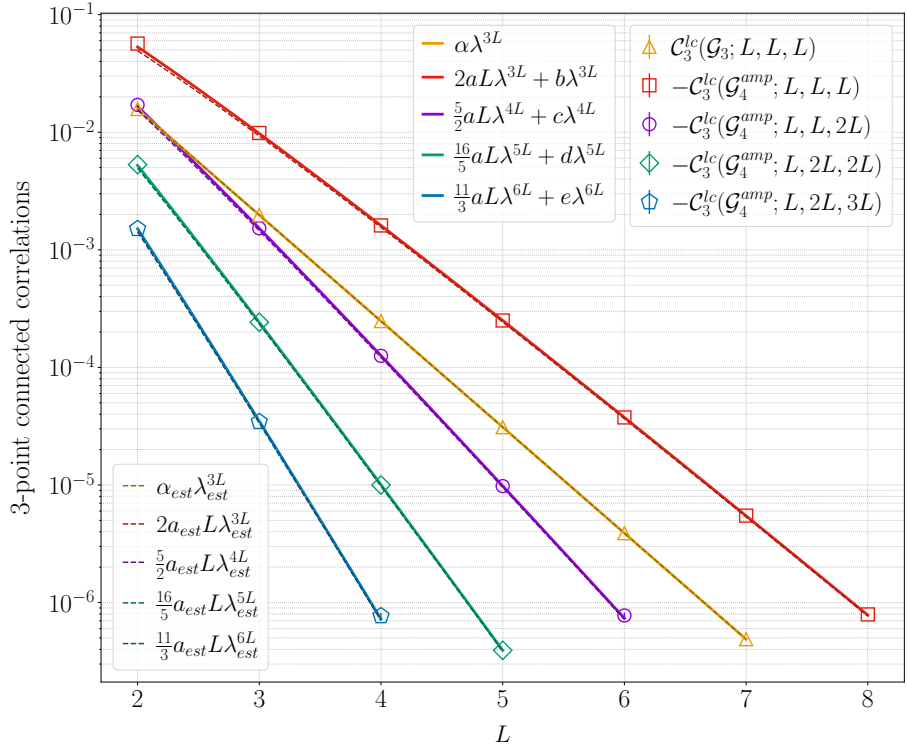

**Fig. S9.** Left:Points correspond to the numerical evaluation of the three-point connected correlation function on the different diagrams of Fig. S2 as a function of the length $L$; full lines correspond to the associated joint fit; dashed lines correspond to the analytical prediction, with the values $\alpha=1$, $\lambda=1/2$ and $a$ extracted from Fig. S7. Right: the evaluated corresponding fit parameters.

$$\mathcal{C}_3^{lc}(\mathcal{G}_3; L, L, L) = \alpha \lambda^{3L} \quad \text{for } L \in [2,7], \quad N_{data} = 1800\,; \qquad [165]$$

$$-\mathcal{C}_3^{lc}(\mathcal{G}_4^{amp}; L, L, L) = 2aL\lambda^{3L} + b\lambda^{3L} \quad \text{for } L \in [4,8], \quad N_{data} = 1800\,; \qquad [166]$$

$$-\mathcal{C}_3^{lc}(\mathcal{G}_4^{amp}; L, L, 2L) = \frac{5}{2}aL\lambda^{4L} + c\lambda^{4L} \quad \text{for } L \in [3,6], \quad N_{data} = 1800\,; \qquad [167]$$

$$-\mathcal{C}_3^{lc}(\mathcal{G}_4^{amp}; L, 2L, 2L) = \frac{16}{5}aL\lambda^{5L} + d\lambda^{5L} \quad \text{for } L \in [3,5], \quad N_{data} = 1800\,; \qquad [168]$$

$$-\mathcal{C}_3^{lc}(\mathcal{G}_4^{amp}; L, 2L, 3L) = \frac{11}{3}aL\lambda^{6L} + e\lambda^{6L} \quad \text{for } L \in [2,4], \quad N_{data} = 1800\,. \qquad [169]$$

In Fig. S10 we show the numerical results for the disconnected function. Here we plotted the results for both $\mathcal{G}_2^{amp}$ and $\mathcal{G}_2$ together with the bare case $\mathcal{G}_1$. In the table, we reported the results of the joint fit.

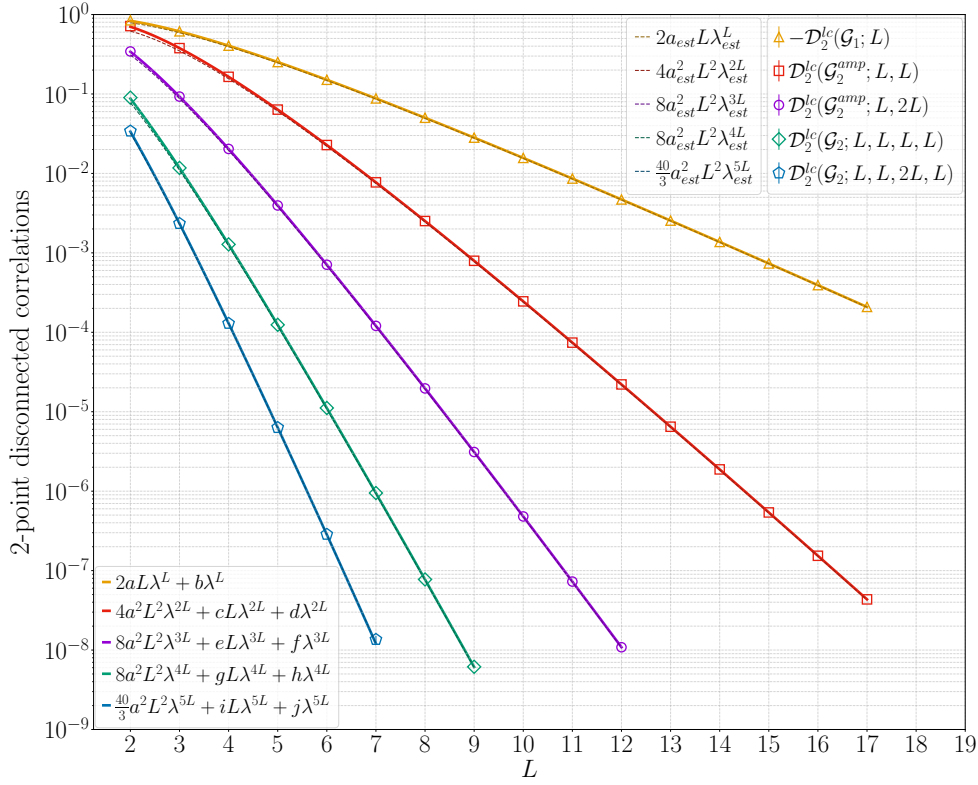

**Fig. S10.** Left: Points correspond to the numerical evaluation of the two-point disconnected correlation function on the different diagrams of Fig. S1 as a function of the length $L$; full lines correspond to the associated joint fit; dashed lines correspond to the analytical prediction, with the values $\alpha = 1$, $\lambda = 1/2$ and $a$ extracted from Fig. S7. Right: the evaluated corresponding fit parameters.

$$-\mathcal{D}_2^{lc}(\mathcal{G}_1; L) = 2aL\lambda^L + b\lambda^L \quad \text{for } L \in [4, 19], \quad N_{data} = 100\,; \tag{170}$$

$$\mathcal{D}_2^{lc}(\mathcal{G}_2^{amp}; L, L) = 4a^2L^2\lambda^{2L} + cL\lambda^{2L} + d\lambda^{2L} \quad \text{for } L \in [3, 17], \quad N_{data} = 200\,; \tag{171}$$

$$\mathcal{D}_2^{lc}(\mathcal{G}_2^{amp}; L, 2L) = 8a^2L^2\lambda^{3L} + eL\lambda^{3L} + f\lambda^{3L} \quad \text{for } L \in [3, 15], \quad N_{data} = 150\,; \tag{172}$$

$$\mathcal{D}_2^{lc}(\mathcal{G}_2; L, L, L, L) = 8a^2L^2\lambda^{4L} + gL\lambda^{4L} + h\lambda^{4L} \quad \text{for } L \in [3, 14], \quad N_{data} = 100\,; \tag{173}$$

$$\mathcal{D}_2^{lc}(\mathcal{G}_2; L, L, 2L, L) = \frac{40}{3}a^2L^2\lambda^{5L} + iL\lambda^{5L} + j\lambda^{5L} \quad \text{for } L \in [3, 9], \quad N_{data} = 200\,. \tag{174}$$



\end{document}